\documentclass[11pt]{article}
\usepackage[top=1in, bottom=1in, left=1in, right=1in]{geometry}
\usepackage[english]{babel}
\usepackage{atbegshi,cite}
\usepackage{amsmath,amssymb,amsbsy,amstext, amsthm, simplewick,textgreek,upgreek,cancel,tensor}
\usepackage{graphicx}
\usepackage{amsfonts}
\usepackage[small]{caption}
\usepackage{upgreek}
\usepackage[titletoc]{appendix}
\usepackage{setspace}
\usepackage{comment}
\usepackage[usenames,dvipsnames,table]{xcolor}
\usepackage[colorlinks=true,urlcolor=blue
,anchorcolor=blue,citecolor=blue,filecolor=blue,linkcolor=blue,menucolor=blue
]{hyperref}

\usepackage{parskip}

\def\pa{\partial}

\newcommand{\newc}{\newcommand}
\newc{\fpi}{f_{\pi}}
\newc{\etap}{\eta^{\prime}}
\newc{\llll}{\langle\lambda\lambda\rangle}
\newc{\FFd}{F^a\tilde F^a}
\newc{\qbar}{{\overline q}}
\newc{\TR}{{\rm Tr}}
\newc{\Kahler}{K\"ahler }
\newc{\Zbb}{{\mathbb Z}}
\newc{\Rt}{{{\mathbb R}^3}}
\newc{\Rf}{{{\mathbb R}^4}}
\newc{\Sth}{{{\mathbb S}^3}}
\newc{\SthSo}{{{\mathbb S}^3\times{\mathbb S}^1}}
\newc{\Stw}{{{\mathbb S}^2}}
\newc{\StwSo}{{{\mathbb S}^2\times{\mathbb S}^1}}
\newc{\So}{{{\mathbb S}^1}}
\newc{\zt}{{{\mathbb Z}_2}}
\newc{\RtSo}{{{\mathbb R}^3\times{\mathbb S}^1}}
\newc{\RfSo}{{{\mathbb R}^4\times{\mathbb S}^1}}
\newc{\scriminus}{{\cal I}^-}
\newc{\scriplus}{{\cal I}^+}
\newc{\mpl}{M_p}
\newc{\Ricci}{\mathcal{R}}
\newc{\bv}{\phi}
\newc{\calU}{{\cal U}}
\newc{\calK}{K}
\newc{\calUi}{{\cal U}^{-1}}
\newc{\calG}{{\cal G}}
\newc{\calI}{{\cal I}}
\newc{\calT}{{\cal T}}
\newc{\calS}{{\cal S}}
\newc{\calM}{{\cal M}}
\newc{\calN}{{\cal N}}
\newc{\calO}{{\cal O}}
\newc{\calQ}{{\cal Q}}
\newc{\calOb}{{\cal O}^\dagger}
\newc{\hphi}{{\hat\phi}}

\newc{\ack}[1]{[{\bf Buzz!: {#1}}]}
\newc{\un}[1]{\underline{#1}}
\newc{\und}[2]{\underbrace{#1}_{(#2)}}
\newc{\prof}[1]{\begin{proof}[\textbf{Proof of }\eqref{#1}]}
\newc{\eprof}{\end{proof}}

\newc{\MK}[1]{\textcolor{blue}{MK: #1}}
\newc{\mk}[1]{\textcolor{red}{MK: #1}}
\newc{\ti}{\tilde}
\newc{\blue}[1]{\textcolor{blue}{#1}}
\newc{\nb}[2]{\textbf{#1.nb~}section: \textbf{#2}}
\newc{\nbb}[1]{\textbf{#1.nb~}}
\newc{\pd}[1]{\textcolor{red}{PD: #1}}

\theoremstyle{plain}
\theoremstyle{plain} 
\theoremstyle{plain} 
\theoremstyle{plain}
\theoremstyle{plain}
\theoremstyle{plain}

\renewcommand{\title}[1]{{\Large\bf\flushleft{#1}}\vspace*{3ex}\\}
\renewcommand{\author}[2]{{\noindent\hspace*{2.5em}\large#1}
                     \footnote{Electronic mail: $\mathtt{#2}$}\\}

\newcommand{\beq}{\begin{equation}}
\newcommand{\eeq}{\end{equation}}

\begin{document}
\begin{titlepage}

\vskip 2.2cm

\begin{center}

{\large \bf Path Integral Factorization and the Gravitational Effective Action}
\vskip 1.4cm

{Patrick Draper$^{(a),}$\footnote{pdraper@illinois.edu},\;Szilard Farkas\footnote{farkas@uchicago.edu},\;and Manthos Karydas$^{(a),}$\footnote{karydas2@illinois.edu}}
\\
\vskip 1cm
{ $^{(a)}$Illinois Center for Advanced Studies of the Universe \&\\Department of Physics, University of Illinois, Urbana, IL 61801}\\
\vspace{0.3cm}
\vskip 4pt

\vskip 1.5cm

\begin{abstract}
We discuss the factorization and continuity properties of fields in the Euclidean gravitational path integral with higher dimension operators constructed from powers of the Riemann tensor. We construct the boundary terms corresponding to the microcanonical ensemble and show that the saddle point approximation to the path integral with a quasilocal energy constraint generally yields a saddle point with discontinuous temperature. This extends a previous result for the Euclidean Schwarzschild-de Sitter geometry in Einstein gravity and shows that it is robust against at least some types of quantum corrections from heavy fields. 
As an  application, we compute the entropy of SdS in $\text{D}=4$ using the BTZ method. Our result matches the entropy calculated using Wald's formula.

\end{abstract}

\end{center}

\vskip 1.0 cm

\end{titlepage}
\setcounter{footnote}{0} 
\setcounter{page}{1}
\setcounter{section}{0} \setcounter{subsection}{0}
\setcounter{subsubsection}{0}
\setcounter{figure}{0}


\tableofcontents

\section{Introduction}
A desirable property of functional integrals is ``factorizability": it should be possible to split the path integral defined on a spacetime manifold $\calM$ into integrals on subregions of $\calM$. For factorization to hold, generally the fields need to respect some continuity properties, so that the classical action functional over subregions is suitably additive. 

As a result of these continuity conditions, the factorized form of the path integral on $\calM$ is not usually a simple product of subregion path integrals, but rather it is given by the integral of such a product over field data defined on the codimension-1 boundaries between the subregions.  Loosely we may think of this decomposition analogously to the insertion of a complete set of states in canonical quantization. Restricting the lower-dimensional path integral so that some data on the interface are fixed corresponds to the insertion of a constraint.

The continuity properties of the fields across a factorization surface follow from the form of the classical action. Consequently there is a close relationship between the factorization properties of the functional integral and the variational problem associated with the semiclassical approximation. Fields that are fixed on the boundary in the classical variational problem are generally required to be continuous across a factorization surface, and thus appear in the lower-dimensional integral over the interface data. Fields that are not included in the lower-dimensional integral over the interface data are not generally continuous. This can have interesting consequences for the semiclassical approximation to factorized path integrals, for fixed factorization surface data or equivalently in the presence of a constraint: the fields at the stationary point may not be continuous.

An example of this phenomenon arises in the Euclidean path integral in 4-dimensional pure Einstein gravity with positive cosmological constant. In the absence of a constraint, there is a completely smooth solution given by the 4-sphere. However, imposing a spherically symmetric constraint on the Brown-York (BY) quasilocal energy~\cite{Brown:1992br}, the relevant solution is given by the Euclidean Schwarzschild de-Sitter (SdS) metric~\cite{Draper:2022xzl, Morvan:2022ybp}. The black hole and cosmological ``horizons" can both be made smooth by a suitable choice of Euclidean time periodicity on either side of the surface where the constraint is imposed, but the choices are different, so the metric is not continuous across the surface. The discontinuity may be encoded as a jump in the lapse function in static slicing. The lapse is not part of the field data that needs to be provided in the variational problem with fixed BY energy, and it is not integrated over in the factorized form of the path integral, so it is perfectly acceptable for it to appear with a discontinuity in the semiclassical solution. At the leading semiclassical order, at least, the on-shell action is independent of the radius at which the constraint is imposed~\cite{Draper:2022xzl}. This resolution of the semiclassical interpretation of SdS also provides support to the idea that localized matter should correspond to a constrained state of the underlying quantum theory of de Sitter space~\cite{Banks:2006rx}.\footnote{See also \cite{Witten:2023xze} and refs therein.} 

One should ask whether discontinuous saddle point geometries obtained in Einstein gravity still make sense beyond the leading semiclassical order. The approach to this question taken here will be to determine whether the factorization properties of more general gravitational Wilsonian effective actions are, in a suitable sense, consistent with the factorization properties of Einstein gravity. Higher-dimension operators can account for fluctuations of short-wavelength degrees of freedom, and so provide a  test of whether the continuity conditions inferred from the Einstein-Hilbert action continue to hold  beyond the leading  order. 

At a technical level, we work with ``$f$(Riemann)" gravity, where the action density is an arbitrary function of the Riemann tensor, in an auxiliary field formulation~\cite{Deruelle:2009zk} where the bulk Lagrangian is second order in derivatives, and we consider ``canonical ensemble" and ``microcanonical ensemble" versions, where boundary terms are chosen so that the semiclassical variational problem corresponds to fixing the induced boundary metric or fixing the boundary energy and momentum fluxes, respectively. The choice of $f$(Riemann) gravity could be viewed as one of technical convenience, since we can borrow the auxiliary field formulation of~\cite{Deruelle:2009zk}. It is not a completely general effective action, as it omits some terms of higher order in covariant derivatives as well as parity odd operators. However, up to dimension four parity-even operators it is general, and in pure gravity through dimension six it is general up to field redefinitions and total derivatives~\cite{Endlich:2017tqa}.\footnote{The dimension four terms $R^2$ and $R_{ab}R^{ab}$ may be removed by a field redefinition $g_{ab}\rightarrow g_{ab}+b R_{ab}+ c\, R g_{ab}$, modulo a shift in $1/G_N$ proportional to the cosmological constant. The third independent dimension four term may be taken to be the Gauss-Bonnet term, a total derivative. The dimension six terms may be categorized as of $f$(Riemann) type, total derivatives, and terms proportional to the vacuum Einstein equation. The latter can again be removed by field redefinitions, up to total derivatives and terms of higher dimension.} To our knowledge there has been limited investigation in the literature of how total derivatives, generated either by integrating out massive fields or by field redefinitions, modify the variational problem needed to implement the semiclassical approximation in effective field theories with boundaries (see, however,~\cite{Jacobson:2013yqa} and refs therein.) We will not take up this problem here, and in any case it is not important for path integral factorization, where any consistent set of boundary conditions can be considered on a factorization surface. (This is analogous to how we may insert either a complete set of position states or a complete set of momentum states in any expectation value in ordinary quantum mechanics.) Instead, we rewrite the general effective Lagrangian of $f$(Riemann) type in the auxiliary field formulation of~\cite{Deruelle:2009zk}, where it is particularly transparent how to implement a consistent Dirichlet-type, or ``canonical ensemble," boundary condition. The auxiliary field formulation has been useful for defining a Brown-York quasi-local stress tensor for curvature squared Lagrangian~\cite{Hohm:2010jc}, with sensible results. The auxiliary field formulation is also essential to pass to a Hamiltonian formulation, and so cleanly facilitates the construction of a second set of consistent ``microcanonical ensemble" boundary conditions. In any event, the class of theories we examine is illustrative of a general procedure which can be applied to other effective actions; for example, higher derivatives may be accommodated by expanding the set of auxiliary fields.

Although we do not know the precise generalization of the Euclidean SdS solution in an arbitrary $f$(Riemann) theory, the presence of two Euclidean horizons at different temperatures implies that a lapse discontinuity is unavoidable. To that end, one of the main results we will obtain is that the microcanonical boundary condition (and associated microcanonical factorization and constraints) can be naturally generalized to theories with higher dimension operators, in such a way that the lapse function is still not part of the boundary data in boundary-value problems or part of the continuous field data in factorization problems. Thus we may infer that a similar ``saddle point of a constrained path integral" interpretation will apply to whatever geometry corrects Euclidean SdS in the semiclassical treatment of more general gravitational effective actions, at least ones that may be cast by field redefinitions to the $f$(Riemann) class. In the simplest case of Einstein gravity plus curvature-squared corrections, SdS remains an exact solution, and we show that the Wald entropy~\cite{Wald:1993nt, Jacobson:1993vj} matches the thermodynamic entropy computed using the BTZ method which is well-adapted to the Hamiltonian formalism~\cite{Banados:1993qp}.

This work is organized as follows. In Section \ref{Sec:Conventions} we set up the notation and conventions for this paper. In Sec. \ref{sec:Factorization Methodology} we give an overview of our approach to path integral factorization. For a given action we explain the importance of the set ${\cal B}_{i}$, which is the set of continuous fields required for the  action to satisfy the additivity property. This is most naturally identified in the purely bulk form of the action. We also point out the close connection between the set ${\cal B}_{i}$ and the boundary conditions needed for a well-posed variational problem. In Sec. \ref{sec:Hamiltonian formulation of f(Riemann) gravity} we review the derivation of the canonical action $I_{\text{ADM}}$ for $f$(Riemann) effective Lagrangians, the path integral factorization of which we  study in Sec. \ref{sec:Path Integral Factorization}. We define the canonical path integral in Eq.	~\eqref{f(Riemann)_ADM_path_integral} and in Appendix \ref{subsec:General Relativity} we exhibit the reduction of the canonical path integral to that of General Relativity for a specific function $f$. We end Sec.~\ref{sec:Hamiltonian formulation of f(Riemann) gravity} with  remarks regarding local symmetries in the canonical form of the action. 
Section \ref{sec:Path Integral Factorization} is split into subsections 
\ref{subsec: Purely bulk form} and \ref{subsec:Microcanonical ensemble} focused on the canonical and microcanonical ensemble respectively. In subsection 
\ref{subsec: Purely bulk form} we find the set ${\cal B}_{i}$ for the canonical ensemble by expressing the action in purely bulk form. Details of the derivation as well as the full form of the purely bulk form are given in Appendix \ref{subsec:Derivation of purely bulk form}. The purely bulk form of the canonical action facilitates defining the microcanonical ensemble, following~\cite{Brown:1992bq}, in subsection \ref{subsec:Microcanonical ensemble}. We also identify the set ${\cal B}_i$ for the microcanonical factorization, which, among other fields, contains the surface energy density $\epsilon_{\text{BY}}$ and surface momentum density $J_{a}$, for which explicit expressions are found in Eq.~\eqref{energy_momentum_surface_densities}. In Section \ref{sec:SdS Entropy} we turn  to applications. We show that SdS in 4 dimensions is a constrained state for the general parity-even effective action with operators up to dimension 4. We calculate the constraints and show that SdS satisfies the equation of motion on either side of the constrained surface. We also compute the entropy from the Euclidean on-shell action and verify that it matches Wald's formula. The details of these computations are presented in Appendix \ref{subsec:Calculation of SdS microcanonical data}. Finally, we end with some discussion and outlook in Sec. \ref{sec:Conclusions}. 

\section{Conventions}
\label{Sec:Conventions}
\begin{figure}[h!]
\begin{center}
\includegraphics[width=2.3in]{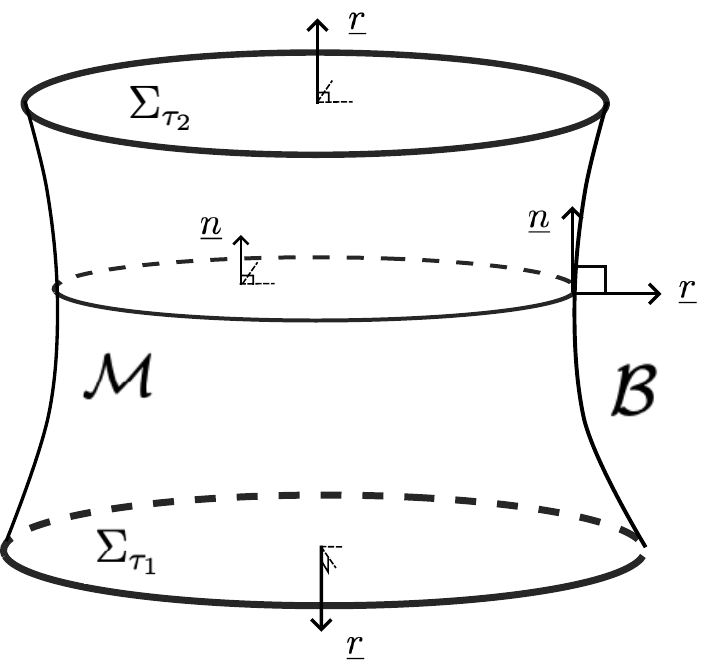}
\caption{The manifold $\calM$ with boundary $\pa \calM= \Sigma_{\tau_{1}}\cup {\cal B}\cup \Sigma_{\tau_{2}}$. The unit normal $\un n$ is future pointing. The unit normal $\un r$ at $\pa\calM$ is defined as outwards pointing. We assume the $\Sigma_{\tau}$ surfaces intersect ${\cal B}$ normally, i.e. $g_{MN} n^{M}r^{N}=0$. The boundary $\cal B$ does not have to be connected, but is drawn that way for simplicity.} 
\label{fig:foliation}
\end{center}
\end{figure}

We consider ambient D-dimensional spacetime $\calM$ with metric $g_{MN}$, cf. Fig. \ref{fig:foliation}. The time coordinate is labeled by $\tau$ and is timelike/spacelike for $\sigma=\mp1$. All our expressions up to Section \ref{sec:SdS Entropy} will hold for general signature, except the explicit path integral expressions in Eqs~\eqref{DSSY_action_2}, \eqref{f(Riemann)_ADM_path_integral}, and~\eqref{microcanonical_path_integral}; it is straightforward to modify them to cover Lorentzian signature. In Section \ref{sec:SdS Entropy} we will work with Euclidean signature only, i.e. $\sigma=+1$. Capital Greek letters $\{M,N,K,\Lambda,\dots \}$ label the coordinates $x^{M}$ of $\calM$ and $\nabla_{M}$ is the metric-compatible covariant derivative on $\calM$.

$\Sigma_{\tau}$  denotes codimension-1 hypersurfaces of constant time $\tau$ with induced metric $h_{\mu\nu}$. Lowercase Greek letters  $\{\mu,\nu,\kappa,\lambda,\dots \}$ (except $\rho$) label coordinates $x^{\mu}$ of $\Sigma_{\tau}$, and $D_{\mu}$ denotes the induced covariant derivative on $\Sigma_{\tau}$, defined as
\beq\label{Induced_derivative}
D_{K}T\indices{^{M_{1}\cdot\cdot M_{p}}_{N_{1}\cdot\cdot N_{q}}}:= \gamma\indices{^{M_{1}}_{M'_{1}}}\cdot\cdot\gamma\indices{^{M_{p}}_{M'_{p}}}\gamma\indices{^{N'_{1}}_{N_{1}}}\cdot\cdot\gamma\indices{^{N'_{q}}_{N_{q}}}\gamma\indices{^{\Sigma}_{K}}\nabla_{\Sigma}\left(\gamma\indices{^{M'_{1}}_{M''_{1}}}\cdot\cdot\gamma\indices{^{N''_{1}}_{N'_{1}}}\cdot\cdot T\indices{^{M''_{1}\cdot\cdot M''_{p}}_{N''_{1}\cdot\cdot N'_{q}}}\right),
\eeq
where $\gamma_{MN}= g_{MN}-\sigma n_{M}n_{N}$ is the projector on $\Sigma_{\tau}$.

${\cal T}$ denotes the codimension-1 factorization surface. It is defined as a surface of constant spatial coordinate $\rho (x^{\mu})$. Throughout this work we assume ${\cal T}$ intersects ${\Sigma_{\tau}}$ normally. When the factorization surface is fixed to a particular $\rho=r$ we denote it as $\calT_{r}$.

${\cal S}_{\tau}= \Sigma_{\tau}\cap {\cal T}$ denotes codimension-2 surfaces defined as the intersection of $\Sigma_{\tau}$ and ${\cal T}$ with induced metric $s_{ab}$. Lowercase Latin letters $\{a,b,c,d,\dots\}$ label coordinates $x^{a}$ of ${\cal S}_{\tau}$ and ${^{(2)}}D_{\mu}$ denotes the codimension-2 induced covariant derivative on ${\cal S}_{\tau}$. In our analysis we will assume ${\calS_{\tau}}$ surfaces are compact and thus drop boundary terms of the form $\int d^{D-2}x \sqrt{s}\,{^{(2)}D_{a}}T^{a}$. For SdS in $\text{D} =4$ the surfaces $\calS_{\tau}$ are 2-spheres.

For the divergence theorem in pseudo-Riemannian manifolds we assume a volume $\calM$ with boundary $\pa\calM={\cup_{i}} \Sigma_{i}$ comprised from surfaces $\Sigma_{i}$ that are solely spacelike/timelike. We define the outward unit normal $r_{M}$ to the boundary $\Sigma_{i}$ with $r_{M}r^{M}=\epsilon_{i}$ and projector $r_{MN}= g_{MN}- \epsilon_{i} r_{M}r_{N}$. The divergence theorem is then
\beq
\label{divergence_pseudo}
\int_{\calM} d^{D}x\sqrt{|g|}\nabla_{M}A^{M}= \int_{\pa{\calM}}d^{D-1}y\,\epsilon_{i}\sqrt{|r|} r_{M}A^{M}\,,
\eeq
where $|r|$ is the determinant of the induced metric on $\pa{\calM}$.
Note that in the case where $\sigma=-1$, the surfaces $\Sigma_{1,2}:=\Sigma_{\tau_{1},\tau_{2}}$ of Figure \ref{fig:foliation} are spacelike with $\epsilon_{1,2}=-1$. For the timelike boundary $\Sigma_{3}:={\cal B}$ we have $\epsilon_{3}=+1$. We will assume ${\cal B}$ and $\Sigma_{\tau}$ intersect normally. We will also drop corner terms, namely boundary total derivatives. 

\section{Factorization Methodology}
\label{sec:Factorization Methodology}
In this section we review 
some aspects of the factorization properties of   gravitational path integrals at leading order in the semiclassical expansion.  
Restricting to the leading order mainly allows us to  be relaxed regarding local functional measures. (For subleading semiclassics there are interesting and subtle issues and ambiguities regarding the path integral measure that will not be addressed here.)

On a manifold $\calM$ of fixed topology the canonical gravitational action takes the form
\beq\label{ADM_action}
I_{\text{ADM}}(\calM)= \int_{\tau_{1}}^{\tau_{2}}d\tau\int_{\Sigma_{\tau}} d^{D-1}x\left({\cal P}_{i}\pa_{\tau}{\cal Q}^{i}\right)- H\left({\cal Q}^{i},{\cal P}_{i},\lambda_{m}\right)\,,
\eeq
where $\lambda_{m}$ are fields that act as Lagrange multipliers. The corresponding path integral is
\beq\label{path_integral_ADM}
Z= \int_{\calM} {{\cal D}{\cal Q}^{i}}\,{\cal D}{\cal P}_{i}\,{\cal D}{\lambda}_{m}\,e^{- I_{\text{ADM}}(\calM)}\,.
\eeq 
Suppose that $\cal M$ is partitioned into two regions $\calM_{1}$, $\calM_{2}$ separated by a surface $\calT$ with $\calM=\calM_{1}\cup \calM_{2}$. Then it is natural to try to write a factorized version of the path integral, 
\beq
\label{path_integral_factorization}
Z= \int_{\calT} D {\cal B}_{i}\,Z(\calM_{1}| {\cal B}_{i})Z( \calM_{2}|{\cal B}_{i})\,.
\eeq
Here the integrand is written as a product of two path integrals over $\calM_{1,2}$ with fixed boundary data ${\cal B}_{i}$ at the interface $\calT$. In the subscript of the integral in (\ref{path_integral_factorization}), 
 $\calT$ is a mnemonic indicating that the  integration variables are fields defined on the boundary $\calT$. The question is which field variables on $\calT$ are included in  ${\cal B}_{i}$?

To answer this question, it is convenient to write the action in a ``purely bulk" form, where any boundary terms (like the GHY term, or the Hamiltonian boundary terms) are rewritten as bulk total derivatives. Suppose that the action has been written in such a form. Then Eq.~\eqref{path_integral_factorization} can hold if the path integrals are over fields satisfying the 
additivity property
\beq
\label{additivity}
I_{\text{ADM}}(\calM)=I^{(1)}_{\text{ADM}}(\calM_{1})+ I^{(2)}_{\text{ADM}}(\calM_{2})\,.
\eeq
The fields ${\cal B}_{i}$ on $\calT$ are given by the restriction to $\calT$ of all fields on $\calM$  whose continuity across $\calT$ is required by the additivity property \eqref{additivity}. (We use ``fields" as a general term that encompasses  fundamental fields, their derivatives and linear combinations, etc.)    
In the purely bulk form of the action, additivity will hold if there are no Dirac delta functions in the action density at $\calT$. Products of discontinuous functions are permissable and such configurations can play an important physical role.\footnote{The simplicity of these conditions motivate but do not require the use of the purely bulk form of the action. If we prefer a form of the action with explicit boundary terms, then the right-hand side of \eqref{additivity}  includes these terms on either side of $\calT$. In this case, the continuity properties required for factorization permit integrable Dirac delta function singularities on $\calT$ in the action density on the left-hand side of \eqref{additivity}, so long as they are captured by the boundary terms on the right-hand side of \eqref{additivity}.}

Let us consider a concrete example. We adopt a coordinate system where $\calT=\calT_r$ lies at a fixed ``radial" coordinate $\rho=r$. (Here and elsewhere ``radial" will be a general term referring to some spatial coordinate, a level surface of which specifies a factorization surface.) Suppose that in a purely bulk form of the action all the terms with radial derivatives have the following form
\beq
\label{action_term_1}
I({\calM})= \int_{\calM} A \pa_{\rho} B\,,
\eeq
where $ A, B$ are fields 
whose exact identity is not needed for now. It is clear that the additivity property holds as long as $B$ is continuous in $\rho$ at $\calT_r$. The field $A$ can be discontinuous in $\rho$. Thus the field $B$ belongs to the set ${\cal B}_{i}$ mentioned before and  must be integrated on $\calT_r$ in the factorized form of the path integral \eqref{path_integral_factorization}, while the field $A$ is integrated over separately in each path integral appearing in the integrand of \eqref{path_integral_factorization}. A standard example of this type is a  canonical path integral in quantum mechanics, where $\rho$ is a time coordinate, $B$ is a particle coordinate, and $A$ is its conjugate momentum.

There is also a close relationship between the continuity properties required for the factorization in \eqref{path_integral_factorization} and the boundary conditions required for a well-defined variational problem in each of the subregions $\calM_{1,2}$. The functional derivatives of the actions  $I^{(1,2)}(\calM_{1,2})$ are well defined for variations in which Dirichlet conditions are imposed on the
${\cal B}_{i}$. If we restrict attention to fields for which the additivity property holds, then we can implement the saddle point approximation on each path integral $Z(\calM_{1,2}| {\cal B}_i)$ separately, with consistent boundary data on $\calT_r$.

If the variational problem is well-defined with Dirichlet conditions on 
the ${\cal B}_{i}$, then we can also apply the saddle approximation to a constrained path integral, namely a path integral where the values of the fields ${\cal B}_{i}$ are held fixed on $\calT_r$ and {\emph {not}} integrated over. This is similar to inserting a projection into a matrix element, or a delta functional into a path integral. The Euclidean SdS solution is precisely a saddle point solution of such
constrained path integral when the $I_{\text{ADM}}$ is the Einstein-Hilbert (EH) action with microcanonical boundary conditions~\cite{Draper:2022xzl}.

In the next sections, we will determine the continuity properties of fields required for factorization of the  gravitational path integral, with action given by a general functional of the Riemann tensor, and with microcanonical boundary terms. We focus on the phase space path integral which is a natural formulation for discussing the microcanonical ensemble.  
As an example application of our results, in Sec. \ref{sec:SdS Entropy} we will show that Euclidean SdS is again a saddle point solution of $\text{D}=4$ $f$(Riemann) gravity theories with microcanonical boundary conditions and operators up to dimension 4, and the on-shell action computes the leading EFT corrections to the entropy.

\section{Canonical action of $f$(Riemann) gravity}
\label{sec:Hamiltonian formulation of f(Riemann) gravity}
In this section we review the derivation of the canonical action $I_{\text{ADM}}$ for $f$(Riemann) theories of gravity, originally presented in~\cite{Deruelle:2009zk}. The main result is given in Eq.~\eqref{ADM_action_1}, which is input to the canonical path integral factorization analysis in Section \ref{sec:Path Integral Factorization}. In this section we keep track both Euclidean/Lorentzian $\sigma=\pm 1$ signature of the time coordinate. 

$f$(Riemann) gravity theories have action of the following form:\footnote{For now we ignore boundary terms in the action. We will introduce them in Eq.~\eqref{DSSY_action_2} after we have defined the auxiliary fields needed for the Hamiltonian description.}
\beq
\label{DSSY_action_1}
S[g_{MN}]= \frac{1}{2}\int_{\calM}d^{D}x \sqrt{|g|}f({\cal R}\indices{^{MN}_{K\Lambda}})\,,
\eeq
where ${\cal R}\indices{^{MN}_{K\Lambda}}$ is the Riemann tensor associated with $g_{MN}$. The Euler-Lagrange equations are
\beq
\label{DSSY_eom_1}
{\cal R}\indices{^{(M}_{K\Lambda P}}\frac{\pa f}{\pa {\cal R}_{N)K\Lambda P}}- 2\nabla_{K}\nabla_{\Lambda}\frac{\pa f}{\pa {\cal R}_{K(MN)\Lambda}}-\frac{1}{2}f g^{MN}=0\,.
\eeq
The action \eqref{DSSY_action_1} generically contains second-order time derivatives which is not suitable for Hamiltonian analysis. To convert the equations of motion \eqref{DSSY_eom_1} into equivalent first order differential equations, one can introduce  auxiliary fields $\varrho_{MNK\Lambda},\varphi^{MNK\Lambda}$ that act as Lagrange multipliers. The action and  path integral  proposed in~\cite{Deruelle:2009zk} is
\begin{align}
\label{DSSY_action_2}
I_{{\cal L}}&= \frac{1}{2}\int_{\calM}d^{D}x \sqrt{|g|}\left[f(\varrho\indices{^{MN}_{K\Lambda}})+ \varphi^{MNK\Lambda}\left({\cal R}_{MNK\Lambda}-{\varrho}_{MNK\Lambda}\right)\right]- {\cal S}_{\text{GHY}}\,,\\\nonumber\\
\label{DSSY_path_integral}
Z&= \int {\cal D} g_{MN}\,{\cal D}\varrho\indices{^{MN}_{K\Lambda}}\,{\cal D}\varphi\indices{^{MN}_{K\Lambda}}\, e^{- I_{{\cal L}}[g_{MN},\varrho\indices{^{MN}_{K\Lambda}},\varphi^{MNK\Lambda}]}\,.
\end{align}
The fields $\varrho_{MNKL},\varphi^{MNKL}$ are assumed to have the same symmetries as the Riemann tensor. The generalized Gibbons-Hawking-York (GHY) term ${\cal S}_{\text{GHY}}$, whose role we will see shortly, is defined as~\cite{SENDOUDA_2011} (see also~\cite{Teimouri:2016ulk})
\beq
\label{Generalized_GHY_term}
{\cal S}_{\text{GHY}}= \int_{\pa \calM}d^{D-1}x\sqrt{|r|}{\cal K}_{MN}{\cal \Psi}^{MN}\,,
\eeq
where  $|r|$ is the determinant of the induced metric on $\pa \calM$ and
\beq
\label{DSSY_K_Psi_GHY}
{\cal K}_{MN}:= r\indices{_{M}^{K}}r\indices{_{N}^{\Sigma}}\nabla_{K}r_{\Sigma}~~,~~{\cal \Psi}^{MN}:= -2 r^{MA}r^{NB}\varphi_{A\Gamma B \Delta}r^{\Gamma}r^{\Delta}\,.
\eeq 
Note that ${\cal K}_{MN}$ is the extrinsic curvature of $\pa\calM$, and by definition $r_{M}$ is the outward pointing unit normal to $\pa \calM$ with $r_{M}r^{M}=\epsilon$. The tensor $r_{MN}= g_{MN}- \epsilon r_{M}r_{N}$ is the projector (see Figure \ref{fig:foliation}) on $\calM$. For timelike outward normal unit vector $\un r$ we have $\epsilon=-1$ and for spacelike $\epsilon=+1$.

At the classical level the auxiliary action \eqref{DSSY_action_2} and $f$(Riemann) action in \eqref{DSSY_action_1} are equivalent, resulting in the same equations of motion for $g_{MN}$. To see this we take the functional variation of the auxiliary action \eqref{DSSY_action_2} 
\beq
\label{bulk_principle_DSSY_action}
\begin{alignedat}{2}
\delta I_{{\cal L}}=& \frac{1}{2}\int_{{\cal M}}d^{D}x \sqrt{|g|}&&\bigg[E^{MN}\delta g_{MN} + \left(\frac{\pa f}{\pa \varrho\indices{^{MN}_{P\Sigma}}} - \varphi^{ABP\Sigma}g_{AM}g_{BN}\right)\delta \varrho\indices{^{MN}_{P\Sigma}} \\&~~~~~&& + \delta \varphi^{MNP\Sigma}\left({\cal R}_{MNP\Sigma}-{\varrho}_{MNP\Sigma}\right) \bigg]\\&
+\epsilon\int_{{\pa \cal M}}d^{D-1}x &&\sqrt{|r|}\left(r_{\Sigma}(\nabla_{P}\varphi^{MPN\Sigma}) \delta g_{MN}- r_{\Sigma} \varphi^{MPN\Sigma}\nabla_{P}\delta g_{MN}\right)- \delta S_{GHY}\,,
\end{alignedat}
\eeq
where 
\beq
\label{metric_variation_DSSY_2}
\begin{split}
E^{MN}&=\frac{1}{2}g^{MN}f(\varrho\indices{^{AB}_{\Gamma\Delta}})+ \frac{1}{2}g^{MN}\varphi^{ABP\Sigma}\left({\cal R}_{ABP\Sigma}-{\varrho}_{ABP\Sigma}\right)\\& 
- 2 \varphi^{(M |ABP}\varrho\indices{^{N)}_{ABP}}
+ \varphi^{ABP (M}{\cal R}\indices{_{ABP}^{N)}}+ 2 \nabla_{A}\nabla_{B}\varphi^{A(MN)B}\,.
\end{split}
\eeq
The equations of motion are
\beq
\label{eom_g_varphi_varrho}
 E^{MN}=0~,~~\varphi\indices{_{MN}^{P\Sigma}}=\frac{\pa f}{\pa \varrho\indices{^{MN}_{P\Sigma}}}~,~~ \varrho_{MNK\Lambda}= {\cal R}_{MNK\Lambda}\,.
\eeq 
If we substitute the second and third equations above into $E^{MN}=0$ we find the $f$(Riemann) metric equations of motion given in \eqref{DSSY_eom_1}. The role of generalized GHY term \eqref{Generalized_GHY_term} is to make the variational problem of the action \eqref{DSSY_action_2} well posed~\cite{SENDOUDA_2011} with the following Dirichlet boundary conditions on $\pa\calM$\footnote{$\delta r^{MN} |_{{\pa \cal M}}$ implies only the induced metric on $\pa\calM$ needs to be fixed. Also note that ${\pa\calM}= \Sigma_{\tau_{1}}\cup {\cal B}\cup \Sigma_{\tau_{2}}$ as depicted in Figure \ref{fig:foliation}.} 
\beq
\label{f(Riemann)_Dirichlet_conditions}
\delta r^{MN} |_{{\pa \cal M}}= \delta \left(r_{K}r_{\Lambda}\varphi^{MK N\Lambda}\right)|_{{\pa\cal M}}=0\,.
\eeq
 The utility of the auxiliary action \eqref{DSSY_action_2} is that it contains only first-order time derivatives, suitable for canonical analysis. We assume a ``time" function $\tau: \calM\to \mathbb{R}$ that defines a foliation of spacetime $\calM$. The surfaces of constant $\tau$ could be timelike ($\sigma=-1$) or spacelike ($\sigma=+1$). The metric is decomposed into an Arnowitt-Deser-Misner (ADM)~\cite{Arnowitt:1962hi} form adapted to the foliation:
\beq
\label{ADM_tau}
ds^2= \sigma N^{2}d\tau^2 + h_{\mu\nu}\left(dx^{\mu}+ N^{\mu}d\tau\right)\left(dx^{\nu}+ N^{\nu}d\tau\right)\,.
\eeq
We define the unit normal form to constant time slices\footnote{Raising the index we get $\un n=N^{-1}\un\pa_{\tau}- N^{-1}N^{\mu}\un\pa_{\mu}$.}
\beq
\label{normal vector_tau}
n_{M}= \sigma N \pa_{M}\tau~~,~~N>0,~~n_{N}n^{N}=\sigma\,.
\eeq
The sign convention for \eqref{normal vector_tau} is such that the dual vector $\un n$ points to increasing  values of $\tau$ i.e. $n^{M}\pa_{M}\tau \ge0$. The time evolution vector $\un \tau= \frac{\pa}{\pa \tau}$, the normal vector $\un n$ and the shift vector $\un N$ are related by
\beq
\label{tau_normal_shift}
\un \tau= N\un n + \un N\,.
\eeq
The codimension-1 projection tensor $\gamma$ on constant time  surfaces is
\beq
\label{Constant_tau_projection}
\gamma_{MN}= g_{MN}-\sigma n_{M}n_{N}\,.
\eeq
A very useful property of the projection tensor that we will repeatedly use is that $\gamma^{\mu\nu}=h^{\mu\nu}$ and $\gamma^{MN}=0$ for $(M,N)\neq (\mu,\nu)$. Here $h^{\mu\nu}$ is the inverse of the induced metric on each time slice $\Sigma_{\tau}$.
The Gauss-Codazzi and Ricci equations are 
\beq
\label{Gauss_Codazzi_Ricci}
\begin{split}
\gamma\indices{^{\Sigma}_{A}}\gamma\indices{^{K}_{B}}\gamma\indices{^{P}_{\Gamma}}\gamma\indices{^{T}_{\Delta}}{\cal R}_{\Sigma K P T}&=\sigma \left(K_{A\Delta}K_{\Gamma B}-K_{A \Gamma}K_{B\Delta}\right)+R_{AB\Gamma\Delta}\,,\\
\gamma\indices{^{\Sigma}_{A}}\gamma\indices{^{K}_{B}}\gamma\indices{^{P}_{\Gamma}}{\cal R}\indices{_{\Sigma KPT}}n^{T}&= D_{A}K\indices{_{B\Gamma}}-D_{B}K\indices{_{A\Gamma}}\,,\\
\gamma\indices{^{\Lambda}_{M}}\gamma\indices{^{K}_{N}}{\cal R}\indices{_{\Lambda P K \Sigma}}n^{P}n^{\Sigma}&=K\indices{_{M}^{\Sigma}}K_{\Sigma N}- \sigma N^{-1}D_{M}D_{N}N- N^{-1}{\cal L}_{N\un n}K_{MN}\,,
\end{split}
\eeq
where $D_{M}$ is the induced covariant derivative on each time slice (see Section \ref{Sec:Conventions}) and $K_{MN}$ is the extrinsic curvature tensor of the $\Sigma_{\tau}$ hypersurface,
\beq
\label{Extrinsic_curvature_tensor_time_slice}
K_{MN}:= \gamma\indices{^{K}_{M}}\gamma\indices{^{\Lambda}_{N}}\nabla_{K}n_{\Lambda}\,.
\eeq

Performing a $(D-1)+1$ decomposition of the fields $\varphi^{MNK\Lambda}, {\cal R}_{MNK\Lambda},\varrho_{MNK\Lambda}$, using the projection tensor $\gamma\indices{^{M}_{N}}$ and  Eqs.~\eqref{Gauss_Codazzi_Ricci}, the action \eqref{DSSY_action_2} becomes
\beq
\label{DSSY_action_3}
I_{{\cal L}}= \int_{\calM}d^{D}x\,{\cal L} - \sigma\int d^{D}x\sqrt{|g|}\nabla_{M}\left(n^{M}K\cdot \Psi\right) - {\cal S}_{\text{GHY}}\,,
\eeq
where\footnote{From \eqref{tau_normal_shift} we have ${\cal L}_{N\un n}\Psi^{\mu\nu}= \dot\Psi^{\mu\nu}- {\cal L}_{\un N}\Psi^{\mu\nu}$.}
\beq
\label{ADM_Lagrangian}
\begin{split}
{\cal L}=N \sqrt{h}\bigg[&\frac{1}{2}f(\varrho_{MNK\Lambda},g_{MN})+  \frac{1}{2}\phi^{\mu\nu\kappa\lambda}\left(\sigma K_{\mu\lambda}K_{\kappa\nu}- \sigma K_{\mu\kappa}K_{\nu\lambda}+ R_{\mu\nu\kappa\lambda} -\rho_{\mu\nu\kappa\lambda}\right)\\&
 + 2\sigma \phi^{\mu\nu\kappa}\left(D_{\mu}K_{\nu\kappa}-D_{\nu}K_{\mu\kappa}-\rho_{\mu\nu\kappa}\right)\\&
 + \sigma \Psi^{\mu\nu}\left((K\cdot K)_{\mu\nu}- \sigma N^{-1}D_{\mu}D_{\nu}N+ K K_{\mu\nu}  - \Omega_{\mu\nu}\right)+ \sigma N^{-1}K_{\mu\nu}{\cal L}_{N\un n}\Psi^{\mu\nu}\bigg]\,,
 \end{split}
\eeq
and the ``spatial" fields are 
\beq
\label{spatial_fields}
\begin{split}
\rho_{\mu\nu\kappa\lambda}&:=\varrho_{\mu\nu\kappa\lambda},~~\rho_{\mu\nu\kappa}:=\varrho_{\mu\nu\kappa P}n^{P},~~ \Omega_{\mu\nu}:= \varrho_{\mu K \nu\Lambda}\,n^{K}n^{\Lambda},\\
\phi^{\mu\nu\kappa\lambda}&:= h^{\mu \mu'}h^{\nu \nu'}h^{\kappa\kappa'}h^{\lambda \lambda'}\varphi_{\mu'\nu'\kappa'\lambda'}\,,~~ \phi^{\mu\nu\kappa}:= h^{\mu \mu'}h^{\nu \nu'}h^{\kappa\kappa'}\varphi_{\mu' \nu' \kappa' P}n^{P},\\
\Psi^{\mu\nu}&:= 2\sigma h^{\mu \mu'}h^{\nu\nu'}n^{K}n^{\Lambda}\varphi_{\mu' K\nu' \Lambda}\,,
\end{split}
\eeq
with $A\cdot B:= A_{\mu\nu}B^{\mu\nu}$, $\left(A\cdot B\right)_{\mu\nu}:= A_{\mu\lambda}B\indices{^{\lambda}_{\nu}}$. Note that the Riemannian symmetries of $\varphi^{MNK\Lambda}$, $\varrho_{MNK\Lambda}$ imply $\phi^{(\mu\nu)\lambda}=\Omega_{[\mu\nu]}= \Psi^{[\mu\nu]}=0$ and $\phi^{\mu\nu\kappa\lambda}$ has the same symmetries as the Riemann tensor. In equation \eqref{ADM_Lagrangian} we assume  $f(\varrho_{MNK\Lambda})$ has been written in terms of the fields $\rho_{\mu\nu\kappa\lambda},\rho_{\mu\nu\kappa},\Omega_{\mu\nu},h_{\mu\nu}$.

There are  two boundary terms appearing in the Lagrangian \eqref{DSSY_action_3}. The first boundary term comes from the Gauss-Codazzi decomposition, and we will refer it as the Gauss-Codazzi boundary term, while the second is the generalized GHY term \eqref{Generalized_GHY_term} introduced at the beginning of this section. Assuming the boundary is of the form $\pa {\calM}= \Sigma_{\tau_{1}}\cup {\cal B}\cup \Sigma_{\tau_{2}}$ as shown in Figure \ref{fig:foliation}, we can use the divergence theorem \eqref{divergence_pseudo} on the Gauss-Codazzi boundary term and check that it exactly cancels the GHY term on $\Sigma_{\tau_{1},\tau_{2}}$. Meanwhile it vanishes on the boundary ${\cal B}$, since we assume the orthogonality relation $g^{MN}n_{M}r_{N}=0$. Thus the only boundary term surviving in the form \eqref{DSSY_action_3} is the GHY term \eqref{Generalized_GHY_term} on the boundary ${\cal B}$, labeled as ${\cal S}^{{\cal B}}_{\text{GHY}}$
\beq
\label{DSSY_action_4}
I_{{\cal L}}= \int_{\calM}d^{D}x\,{\cal L} - {\cal S}^{{\cal B}}_{\text{GHY}}\,,
\eeq
where ${\cal L}$ is given in \eqref{ADM_Lagrangian}. The GHY term expressed in terms of the spatial fields is
\beq
\label{GHY_canonical}
\begin{split}
{\cal S}^{\cal B}_{\text{GHY}}&=\int_{{\cal B}}d^{D-1}x\sqrt{|r|}\bigg[-2 k_{\mu\nu}\phi^{\mu\gamma\nu\delta}r_{\gamma}r_{\delta}- 4\sigma r^{\beta}K_{\beta\alpha}\phi^{\alpha\gamma\delta}r_{\gamma}r_{\delta} - (D_{\un r}\ln N)(r_{\alpha}r_{\beta}\Psi^{\alpha\beta})\bigg]\,,
\end{split}
\eeq
where $\sqrt{|r|}=N \sqrt{s}$, $D_{\un r}:= r^{\mu}D_{\mu}$ and $r_{\mu}$ is the unit normal one form at ${\cal B}$ which is also tangent to $\Sigma_{\tau}$.

The Lagrangian \eqref{ADM_Lagrangian} is first order in time derivatives since the constant time extrinsic curvature tensor is\footnote{\label{footnote1}For all ambient components $K_{MN}$ we find from \eqref{Extrinsic_curvature_tensor_time_slice} that $K_{MN}=\frac{1}{2N}{\cal L}_{N\un n}\gamma_{MN}=\frac{1}{2N}\dot \gamma_{MN}- \frac{1}{2N}{\cal L}_{\un N}\gamma_{MN}$.} 
\beq
\label{Extrinsic_Lie_der}
K_{\mu\nu}=\frac{1}{2N}(\pa_{\tau}h_{\mu\nu}) - \frac{1}{2N}{\cal L}_{\un N}h_{\mu\nu}\,,
\eeq
The non-vanishing canonical momentum densities are
\beq
\label{Hamiltons_first_equations}
p^{\mu\nu}:= \frac{\delta {\cal L}}{\delta(\pa_{\tau} h_{\mu\nu})}~~,~~\Pi_{\mu\nu}:= \frac{\delta {\cal L}}{\delta (\pa_{\tau}\Psi^{\mu\nu})}\,,
\eeq
and can be calculated from the action \eqref{DSSY_action_4}. We find 
\beq
\label{momenta}
\begin{split}
p^{\mu\nu}= \sigma\sqrt{h}\bigg[& \phi^{\mu\alpha\beta\nu}K_{\alpha\beta} - 2 N^{-1}D_{\beta}\left(N\phi^{\beta (\mu\nu)}\right)+ \left(\Psi\cdot K\right)^{(\mu\nu)}+ \frac{1}{2} \left(K\cdot \Psi\right)h^{\mu\nu}\\&+ \frac{1}{2} K \Psi^{\mu\nu}+ \frac{1}{2}N^{-1}{\cal L}_{N\un n}\Psi^{\mu\nu}\bigg]\,,\\
\Pi_{\mu\nu}= \sqrt{h}\sigma &K_{\mu\nu}\,.
\end{split}
\eeq
The Hamiltonian integral is 
\beq\label{Hamiltonian_integral}
\int_{\tau_{1}}^{\tau_{2}} d\tau H\,=\int_{\tau_{1}}^{\tau_{2}} d\tau \int_{\Sigma_{\tau}}d^{D-1}x\left( p^{\mu\nu}\pa_{\tau} h_{\mu\nu}+ \Pi_{\mu\nu}\pa_{\tau} \Psi^{\mu\nu} {\cal }\right)- I_{\cal L}\,,
\eeq
where $I_{{\cal L}}$ is the action given in \eqref{DSSY_action_4}. We find
\beq\label{Hamiltonian_0}
\int_{\tau_{1}}^{\tau_{2}} d\tau H= \int_{\tau_{1}}^{\tau_{2}} d\tau \int_{\Sigma_{\tau}}d^{D-1}x\bigg[N^{\mu}{\cal C}_{\mu} + N{\cal C}+ \sqrt{h}D_{\mu}V^{\mu}\bigg]+ {\cal S}^{{\cal B}}_{\text{GHY}}\,,
\eeq
where ${\cal S}_{\text{GHY}}$ is given in \eqref{GHY_canonical} and 
\beq
\label{boundary_terms_V_mu}
V^{\mu}=2\frac{p^{\mu\nu}}{\sqrt{h}}N_{\nu}- 2\frac{\Pi_{\alpha\beta}}{\sqrt{h}}\Psi^{\beta\mu}N^{\alpha}+ \Psi^{\mu\beta}D_{\beta}N- ND_{\beta}\Psi^{\beta\mu}\,,
\eeq
\beq
\label{boundary_terms_lapse_shift_constraints}
\begin{split}
{\cal C}_{\mu}&=2\sqrt{h}D_{\nu}\left(\frac{\Pi\indices{_{\mu}_{\lambda}}}{\sqrt{h}}\Psi^{\lambda \nu}\right) -2 \sqrt{h}h_{\mu\sigma}D_{\nu}\left(\frac{p^{\sigma\nu}}{\sqrt{h}}\right)+ \sqrt{h}\left(\frac{\Pi_{\alpha\beta}}{\sqrt{h}}D_{\mu}\Psi^{\alpha\beta}\right)\,,\\
{\cal C}&=\sqrt{h} \left(D_{\mu}D_{\nu}\Psi^{\mu\nu}- \frac{1}{2}f(\varrho\indices{^{MN}_{K\Lambda}})+\sigma \Psi\cdot \Omega\right)+ \frac{1}{\sqrt{h}}\left(2\sigma p\cdot\Pi - \sigma\Psi\cdot\Pi\cdot\Pi- \sigma \Pi (\Psi\cdot\Pi)\right)\\&~~
-\frac{\sqrt{h}}{2}\phi^{\mu\nu\kappa\lambda}\left(\frac{2\sigma}{h}\Pi_{\mu\lambda}\Pi_{\nu\kappa}+ R_{\mu\nu\kappa\lambda}-\rho_{\mu\nu\kappa\lambda}\right) + 2\sigma \sqrt{h}\phi^{\mu\nu\kappa}\left(\rho_{\mu\nu\kappa}-2\sigma D_{\mu}\left(\frac{\Pi_{\nu\kappa}}{\sqrt{h}}\right)\right)\,.
\end{split}
\eeq
From equations \eqref{DSSY_action_4}, \eqref{Hamiltonian_integral} and  \eqref{Hamiltonian_0}  we can express the action \eqref{DSSY_action_2} in \emph{canonical form}
\beq
\label{ADM_action_1}
I_{\text{ADM}}= \int_{\tau_{1}}^{\tau_{2}}d\tau\int_{\Sigma_{\tau}}d^{D-1}x\left(p^{\mu\nu}\pa_{\tau} h_{\mu\nu}+ \Pi_{\mu\nu}\pa_{\tau}\Psi^{\mu\nu}- N{\cal C}- N^{\mu}{\cal C}_{\mu}\right) + B_{\un N}+ B_{N}\,,
\eeq
where $B_{\un N}$, $B_{N}$ are \footnote{We use the notation $B_{N}$, $B_{\un N}$ since they reduce to the General Relativity analogs~\cite{Regge:1974zd} (see Appendix \ref{subsec:General Relativity}).} 
\begin{align}
\label{f(Riemann)_boundary_terms_Hamiltonian}
B_{\un N}&= \int_{\tau_{1}}^{\tau_{2}}d\tau\int_{\calS_{\tau}}d^{D-2}x\sqrt{s}\, N^{\mu}\left(2 \frac{\Pi\indices{_{\mu}_{\lambda}}}{\sqrt{h}}\Psi^{\lambda\nu}- 2h_{\mu\sigma}\frac{p^{\sigma\nu}}{\sqrt{h}}\right)r_{\nu}\,,\\
\label{f(Riemann)_boundary_terms_Hamiltonian_2}
B_{N}&= \int_{\tau_{1}}^{\tau_{2}}d\tau\int_{\calS_{\tau}}d^{D-2}x\sqrt{s}\, N\left(D_{\lambda}\Psi^{\lambda\nu}- \Psi^{\nu\lambda}D_{\lambda}\ln N\right)r_{\nu} - {\cal S}^{{\cal B}}_{\text{GHY}}\,,
\end{align}
and ${\cal C},{\cal C}_{\mu}$ are given in \eqref{boundary_terms_lapse_shift_constraints} and ${\cal S}^{{\cal B}}_{\text{GHY}}$ in \eqref{GHY_canonical}. The corresponding canonical path integral is
\beq
\label{f(Riemann)_ADM_path_integral}
Z= \int {\cal D} N{\cal D}N^{\mu}{\cal D}h_{\mu\nu}{\cal D}p^{\mu\nu}{\cal D}\Psi^{\mu\nu}{\cal D}\Pi_{\mu\nu}{\cal D}[\Omega_{\mu\nu},\rho_{\mu\nu\kappa\lambda},\rho_{\mu\nu\kappa}\phi^{\mu\nu\kappa\lambda}\phi^{\mu\nu\kappa}]\, e^{- I_{\text{ADM}}}\,,
\eeq
where $I_{\text{ADM}}$ is given in \eqref{ADM_action_1}. 
The path integral \eqref{f(Riemann)_ADM_path_integral} corresponds to the canonical ensemble~\cite{Brown:1992br} of $f$(Riemann) gravity, since part of the boundary data for a well-defined variational problem requires fixing the induced metric on $\pa\calM$, as can be verified from Eq. \eqref{f(Riemann)_Dirichlet_conditions}. We will see this explicitly in the next section.

In Appendix \ref{subsec:General Relativity} we reduce the path integral \eqref{f(Riemann)_ADM_path_integral} to the familiar one of General Relativity when $f^{\text{GR}}(\varrho\indices{^{MN}_{K\Lambda}})= -\frac{\sigma}{8\pi G_{N}}\left(g^{MP}g^{N\Sigma}\varrho_{MNP\Sigma}-2\Lambda\right)$.

\subsection{Local Symmetries}
\label{subsec:Local Symmetries} 

Our analysis is suitable for the evaluation of path integrals, with fixed data on a factorization surface, in the semiclassical approximation. In going beyond the leading order, one must identify the gauge symmetries of the action and fix them. We will not do so here, but we make some brief remarks about this procedure. 

In the canonical formalism of GR, the constraints are first class and depend on the canonical variables only. They generate the symmetries of the canonical variables and their Poisson algebra also determines the transformation rules of the lapse and the shifts~\cite{Teitelboim:1972vw}. We have written the $f$(Riemann) path integral in the Hamiltonian formalism retaining auxiliary fields, which are extra variables that are not canonical. The status of local symmetries of the off-shell action is more obscure in this formulation of $f$(Riemann) than it is in ordinary GR.

By solving the equation of motion of the canonical momenta, one can fall back on the Lagrangian formulation~(\ref{DSSY_action_2}) and write down the path integral with a manifestly diffeomorphism-invariant action. Another option can be the elimination of the auxiliary fields by solving some of their equations of motion. This way one gets an action that depends only on the lapse and the shifts in addition to the canonical variables. The gauge symmetries of this action can be identified as in canonical GR. 

In either case, the auxiliary variables are of great use for reading off continuity properties in the path integral factorization, and the Hamiltonian formalism is useful -- although not essential -- for working out the microcanonical boundary terms. The possibility of evaluating the path integral in the auxiliary field formulation beyond the leading saddle point approximation will be investigated in future work.

\section{$f$(Riemann) Path Integral Factorization}
\label{sec:Path Integral Factorization}
In this section we find the continuity conditions for the path integral factorization of $f$(Riemann) gravity theories in both canonical and microcanonical ensemble. 
For this purpose we need to isolate all radial derivatives appearing in the purely bulk form of the canonical action \eqref{ADM_action_1}, which corresponds to the canonical ensemble. (Again we stress that ``radial" is used as a general term describing a spatial coordinate, constant values of which will label factorization surfaces. By a change of coordinatization we can work with any factorization surface orthogonal to the temporal foliation.)  To achieve this we employ another ADM type decomposition of the induced metric and the rest of spatial fields on $\Sigma_{\tau}$ along a foliation of constant $\rho(x^{\mu})$ surfaces. This allows us to implement a Gauss-Codazzi type decomposition of the canonical action and convert it to purely bulk form such that the  $\rho$-derivatives appear explicitly. The purely bulk form is also important from another perspective; it will allow us to define the microcanonical ensemble by adding the necessary boundary terms to the canonical action.

\subsection{Purely bulk form of $I_{\text{ADM}}$}
\label{subsec: Purely bulk form}
In Section \ref{sec:Hamiltonian formulation of f(Riemann) gravity} we derived the canonical action $I_{\text{ADM}}$ given in Eq.~\eqref{ADM_action_1} for the $f$(Riemann) Lagrangian \eqref{DSSY_action_2}. 
The canonical action $I_{\text{ADM}}$ is obviously not in purely bulk form due to the existence of the boundary terms $B_{N}$ and $B_{\un N}$ given in Eqs.~\eqref{f(Riemann)_boundary_terms_Hamiltonian}-\eqref{f(Riemann)_boundary_terms_Hamiltonian_2}. Both these boundary terms can be cancelled by surface terms produced from some of the bulk terms after integration by parts.

To isolate these terms we consider a foliation of the time slices $\Sigma_{\tau}$ by surfaces of constant radial coordinate $\rho (x^{\mu})$. We then employ another ADM decomposition of the induced metric $h_{\mu\nu}$ as follows
\beq
\label{ADM_cod_2_purely_bulk}
\begin{split}
ds^{2}|_{\Sigma_{\tau}}:&=h_{\mu\nu}dx^{\mu}dx^{\nu}= P^{2}d\rho^{2}+ s_{ab}(dx^{a}+ \beta^{a}d\rho)(dx^{b}+ \beta^{b}d\rho)\,,
\end{split}
\eeq
where $P:= \sqrt{h_{\rho\rho}}$ is the radial lapse, $\un\beta$ is the radial shift vector and $s_{ab}$ is the induced metric on the codimension-2 surface $\calS_{\tau}= \Sigma_{\tau}\cap \calT$. The Latin indices label the coordinates $x^{a}$ on ${\cal S}_{\tau}$. The unit one form $r_{\mu}$ and dual vector $\un r$ normal to $\rho$- surfaces on $\Sigma_{\tau}$ are
\beq
\label{radial_unit_form}
r= P d\rho\,,~~~\un r= P^{-1}\un\pa_{\rho}- P^{-1}\un \beta\,,
\eeq
where $\un\beta=\beta^{a}\un\pa_{a}$ and $r_{\mu}r^{\mu}=1$. The ${\cal S}_{\tau}$ - projector $r_{\mu\nu}=h_{\mu\nu}-r_{\mu}r_{\nu}$ allows us to define the \emph{radial extrinsic curvature tensor} 
\beq
\label{radial_extrinsic_tensor}
k_{\mu\nu}=r\indices{^{\sigma}_{\mu}}r\indices{^{\lambda}_{\nu}}D_{\sigma}r_{\lambda}\,.
\eeq
One can show that $k_{\mu\nu}$ contains radial derivatives of $s_{ab}$ only.
The Riemann tensor on $\Sigma_{\tau}$ has the familiar Gauss-Codazzi decomposition (analogous to Eqs.~\eqref{Gauss_Codazzi_Ricci})
\beq
\label{Gauss_Codazzi_Ricci_cod_2_v2}
\begin{split}
r\indices{^{\sigma}_{\alpha}}r\indices{^{\kappa}_{\beta}}r\indices{^{\rho}_{\gamma}}r\indices{^{\tau}_{\delta}}{ R}_{\sigma \kappa \rho \tau}&= \left(k_{\alpha\delta}k_{\gamma \beta}-k_{\alpha \gamma}k_{\beta\delta}\right)+ {^{(2)}}R_{\alpha\beta\gamma\delta}\\
r\indices{^{\sigma}_{\alpha}}r\indices{^{\kappa}_{\beta}}r\indices{^{\rho}_{\gamma}}{ R}\indices{_{\sigma \kappa \rho \tau}}r^{\tau}&= {^{(2)}}D_{\alpha}k\indices{_{\beta\gamma}}-{^{(2)}}D_{\beta}k\indices{_{\alpha\gamma}}\\
r\indices{^{\lambda}_{\mu}}r\indices{^{\kappa}_{\nu}}{ R}\indices{_{\lambda \rho \kappa \sigma}}r^{\rho}r^{\sigma}&= -k\indices{_{\mu}^{\sigma}}k_{\sigma \nu}+ {^{(2)}}D_{\mu}a_{\nu} - a_{\mu}a_{\nu} - D_{\un r}k_{\mu\nu}- 2r_{(\mu}k_{\nu)\lambda}a^{\lambda}\,,
\end{split}
\eeq
where ${^{(2)}}D_{\mu}$ and ${^{(2)}}R_{\alpha\beta\gamma\delta}$ denote the induced covariant derivative and Riemann tensor on ${\cal S}_{\tau}$ and $a_{\mu}:= r^{\sigma}D_{\sigma}r_{\mu}$ is the \emph{radial acceleration}. One can check that $a$ has the general expression
\beq
    a_{\mu}= -{^{(2)}}D_{\mu}\ln P,
\eeq 
and thus it does not contain any radial derivatives. From the Gauss-Codazzi equations in Eqs.~\eqref{Gauss_Codazzi_Ricci_cod_2_v2} we can  spot where the radial derivatives are hiding in the Riemann tensor components $R_{\mu\nu\kappa\lambda}$. Most of them are contained inside the radial extrinsic curvature tensor $k_{\mu\nu}$ which contain radial derivatives of $s_{ab}$. There is also a radial derivative of the extrinsic curvature tensor itself that appears as the fourth term in the third equation of \eqref{Gauss_Codazzi_Ricci_cod_2_v2}. To make other radial derivatives in the canonical action visible we first use the radial projector $r_{\mu\nu}$ to decompose (some of) the canonical fields into tensors tangent to $\calS_{\tau}$ similar to the ADM decomposition of the metric field $h_{\mu\nu}$ in \eqref{ADM_cod_2_purely_bulk}. We find
\beq
\label{canonical_fields_radial_projection}
\begin{split}
N^{\mu}&= N^{\mu}_{|}+ r^{\mu}N_{0}\\
\Pi_{\mu\nu}&= \Pi^{|}_{\mu\nu}+ 2r_{(\mu}\Pi^{|}_{\nu)}+ r_{\mu}r_{\nu}\Pi_{0}\\
\Psi^{\mu\nu}&= \Psi^{\mu\nu}_{|}+ 2 r^{(\mu}\Psi^{\nu)}_{|} + r^{\mu}r^{\nu}\Psi_{0}\\
\phi^{\mu\nu\kappa}&= \phi_{|}^{\mu\nu\kappa}+ 2r^{[\nu}\phi^{\mu]\kappa}_{2,|}+ \left(\phi^{\mu\nu}_{1,|}+ 2 r^{[\nu}\phi^{\mu]}_{|}\right)r^{\kappa}\\
\phi^{\mu\nu\kappa\lambda}&= \theta^{\mu\nu\kappa\lambda}_{|}+ 2\theta_{|}^{\mu\nu[\kappa}r^{\lambda]}+ 2\theta_{|}^{\kappa\lambda[\mu}r^{\nu]}+ 2r^{\nu}\theta_{|}^{\mu[\kappa}r^{\lambda]}+ 2 r^{\mu}\theta^{\nu[\lambda}_{|}r^{\kappa]}\,,
\end{split}
\eeq
where we defined the fields
\beq
\label{canonical_fields_radial_projection_definitions}
\begin{split}
N_{|}^{\mu}&:= r\indices{^{\mu}_{\mu'}}N^{\mu'}~~,~~N_{0}:= r_{\lambda}N^{\lambda}\,, \\
\Pi^{|}_{\mu\nu}&:= r\indices{^{\mu'}_{\mu}}r\indices{^{\nu'}_{\nu}}\Pi_{\mu'\nu'}~~,~~ \Pi^{|}_{\mu}:= r\indices{^{\mu'}_{\mu}}\Pi_{\mu'\lambda}r^{\lambda}~~,~~ \Pi_{0}:= \Pi_{\alpha\beta}r^{\alpha}r^{\beta}\,,\\
\Psi_{|}^{\mu\nu}&:=r\indices{^{\mu}_{\mu'}}r\indices{^{\nu}_{\nu'}}\Psi^{\mu'\nu'}~~,~~ \Psi_{|}^{\mu}:= r\indices{^{\mu}_{\mu'}}\Psi^{\mu'\lambda}r_{\lambda}~~,~~\Psi_{0}:= \Psi^{\alpha\beta}r_{\alpha}r_{\beta}\,,\\
\theta^{\mu\nu\kappa\lambda}_{|}&:= r\indices{^{\mu}_{\mu'}}r\indices{^{\nu}_{\nu'}}r\indices{^{\kappa}_{\kappa'}}r\indices{^{\lambda}_{\lambda'}}\phi^{\mu'\nu'\kappa'\lambda'}~,~\theta_{|}^{\mu\nu\kappa}:= r\indices{^{\mu}_{\mu'}}r\indices{^{\nu}_{\nu'}}r\indices{^{\kappa}_{\kappa'}}\phi^{\mu'\nu'\kappa'\lambda}r_{\lambda}~,~\theta^{\mu\nu}_{|}:= r\indices{^{\mu}_{\mu'}}r\indices{^{\nu}_{\nu'}}\phi^{\mu'\alpha\nu'\beta}r_{\alpha}r_{\beta}\,, \\
\phi^{\mu\nu\kappa}_{|}&:=r\indices{^{\mu}_{\mu'}}r\indices{^{\nu}_{\nu'}}r\indices{^{\kappa}_{\kappa'}}\phi^{\mu'\nu'\kappa'},\phi^{\mu\nu}_{1,|}:=r\indices{^{\mu}_{\mu'}}r\indices{^{\nu}_{\nu'}}\phi^{\mu'\nu'\lambda}r_{\lambda},\,\phi^{\mu\nu}_{2,|}:= r\indices{^{\mu}_{\mu'}}r\indices{^{\nu}_{\nu'}}\phi^{\mu'\lambda\nu'}r_{\lambda},~\phi_{|}^{\mu}:= r\indices{^{\mu}_{\mu'}}\phi^{\mu'\alpha\beta}r_{\alpha}r_{\beta}\,.
\end{split}
\eeq
The symbol $|$ above serves as a reminder that the field is tangent to $\calS_{\tau}$, e.g. $r_{\lambda}\Psi_{|}^{\lambda\nu}=r_{\lambda}\Psi_{|}^{\mu\lambda}=0$ and similarly for the other fields. The symmetries of the spatial fields are inherited to the projected fields above, e.g. $\phi^{(\mu\nu)}_{1,|}=0$ since $\phi^{(\mu\nu)\kappa}=0$. All other spatial fields that do not appear above need not be decomposed since they do not participate with terms that have radial derivatives in the canonical action. 

The Gauss-Codazzi equations in Eqs.~\eqref{Gauss_Codazzi_Ricci_cod_2_v2} together with the projected fields in Eqs.~\eqref{canonical_fields_radial_projection_definitions} can be used to express the canonical action \eqref{ADM_action_1} in purely bulk form. For economic purposes we present a shortened version of the purely bulk form up to a function ${\cal G}(\text{fields})$, namely
\beq
\label{Purely_bulk_form_Canonical_action_final_short}
\begin{split}
&I_{\text{ADM}}=\int_{\tau_{1}}^{\tau_{2}}d\tau \int_{\Sigma_{\tau}}d^{D-1}x \left(p^{\mu\nu}\pa_{\tau}h_{\mu\nu}+ \Pi_{\mu\nu}\pa_{\tau}\Psi^{\mu\nu}\right)\\&
+\int_{\tau_{1}}^{\tau_{2}}d\tau \int_{\Sigma_{\tau}}d^{D-1}x \sqrt{h}\blue{(D_{\un r} N)}\bigg[2\,{^{(2)}}D_{\lambda}\Psi^{\lambda}_{|}-\blue{k_{\alpha\beta}}\Psi^{\alpha\beta}_{|} + \blue{k} \Psi_{0}- 2a_{\lambda}\Psi^{\lambda}_{|}+ \blue{D_{\un r}\Psi_{0}} + 2\blue{k_{\alpha\beta}}\theta^{\alpha\beta}_{|} + 4\phi^{\lambda}_{|}\frac{\Pi^{|}_{\lambda}}{\sqrt{h}}\bigg]\\&
+\int_{\tau_{1}}^{\tau_{2}}d\tau \int_{\Sigma_{\tau}}d^{D-1}x\sqrt{h}\bigg[(\blue{D_{\un r}N^{\mu}_{|}})\bigg(2\frac{\Pi_{\mu\lambda}}{\sqrt{h}}\Psi^{\lambda\nu}r_{\nu}- 2 \frac{h_{\mu\sigma}p^{\sigma\nu}r_{\nu}}{\sqrt{h}}\bigg) + \blue{D_{\un r}N_{0}}\bigg(\frac{2r^{\mu}\Pi_{\mu\lambda}}{\sqrt{h}}\Psi^{\lambda\nu}r_{\nu}- \frac{2r_{\mu}p^{\mu\nu}r_{\nu}}{\sqrt{h}}\bigg) \bigg]\\&
- 4\int_{\tau_{1}}^{\tau_{2}}d\tau \int_{\Sigma_{\tau}}d^{D-1}x\sqrt{h}N\phi^{\nu\kappa}_{2,|}\blue{D_{\un r}\left(\frac{\Pi^{|}_{\nu\kappa}}{\sqrt{h}}\right)}+ 2\int_{\tau_{1}}^{\tau_{2}}d\tau \int_{\Sigma_{\tau}}d^{D-1}x\sqrt{h}N\blue{k_{\mu\nu}D_{\un r}\theta^{\mu\nu}_{|}}\\&
 + 4\int_{\tau_{1}}^{\tau_{2}}d\tau \int_{\Sigma_{\tau}}d^{D-1}x\sqrt{h}N (\blue{D_{\un r}\phi^{\mu}_{|}})\frac{\Pi^{|}_{\mu}}{\sqrt{h}}\\&
-\int_{\tau_{1}}^{\tau_{2}}d\tau \int_{\Sigma_{\tau}}d^{D-1}x\sqrt{h}N_{0}\left(\frac{\Pi^{|}_{\alpha\beta}}{\sqrt{h}}\blue{D_{\un r}\Psi^{\alpha\beta}_{|}}+ 2\frac{\Pi^{|}_{\beta}}{\sqrt{h}}\blue{D_{\un r}\Psi^{\beta}_{|}} + \frac{\Pi_{0}}{\sqrt{h}}\blue{D_{\un r}\Psi_{0}}\right)\\&
+ \int_{\tau_{1}}^{\tau_{2}}d\tau \int_{\Sigma_{\tau}}d^{D-1}x\sqrt{h}\,{\cal G}\left(\blue{k_{\mu\nu}},a_{\mu}, \text{other field combinations with no radial derivatives}\right)\,.
\end{split}
\eeq 
The detailed derivation of \eqref{Purely_bulk_form_Canonical_action_final_short} is presented in Appendix \ref{subsec:Derivation of purely bulk form} where the exact form of the function ${\cal G}(\text{fields})$ is also given (see Eq.~\eqref{Factorization_ADM_action_purely_bulk_complete_form}). The blue terms in \eqref{Purely_bulk_form_Canonical_action_final_short}  are all the terms that contain radial derivatives. 

Let us make some important remarks. First, note that the blue terms in \eqref{Purely_bulk_form_Canonical_action_final_short} contain radial (spatial) covariant derivatives and not partial derivative of the form $A\pa_{\rho}B$. One could wonder whether there are more radial derivatives in the Christoffel symbols $\gamma\indices{^{\lambda}_{\mu\nu}}$ of the spatial metric $h_{\mu\nu}$. From the radial ADM decomposition in Eq.~\eqref{ADM_cod_2_purely_bulk} one can show that the Christoffel symbols needed contain only (partial) radial derivatives of the induced metric $\pa_{\rho}s_{ab}$. The reason is that the radial covariant derivatives appearing in \eqref{Purely_bulk_form_Canonical_action_final_short} act only on tensors tangential to $\calS_{\tau}$, which the subscript symbol $|$ signifies, as explained below Eqs.~\eqref{canonical_fields_radial_projection_definitions}.\footnote{For $D_{\un r}$ acting on tangential tensors on $\calS_{\tau}$ the Christoffels $\gamma\indices{^{\rho}_{\rho\rho}}$, $\gamma\indices{^{a}_{\rho\rho}}$ are not used. These are the only components that contain radial derivatives of $h_{\rho\rho}$.}

Second, from the purely bulk form of the canonical action we can easily read off the continuity conditions required for factorization as described in Section \ref{sec:Factorization Methodology}. Since our analysis assumes the factorization surface ${\cal T}$ intersects the time slices $\Sigma_{\tau}$ orthogonally, we have $N_{0}|_{{\cal T}}=0$, and the purely bulk from in  Eq.~\eqref{Purely_bulk_form_Canonical_action_final_short} implies  that the following fields\footnote{Without loss of generality in Eq.~\eqref{Continuity_fields_across_T_canonical_action} we replaced lowercase greek $\mu,\nu,\dots$ indices with latin $a,b,\dots$ on $\calS_{\tau}$.} need to be continuous in $\rho$ across ${\cal T}$
\beq
\label{Continuity_fields_across_T_canonical_action}
{\cal B}_{i}^{\text{canonical}}= \{N,~ N^{a},~s_{ab},~ \theta^{ab}_{|},~\phi^{a}_{|},~\Psi_{0}\,,~\text{and }~ \Pi^{|}_{ab}/\sqrt{h}\,\}\quad\mbox{at}~\calT\,.
\eeq
For the variational problem of the canonical action $I_{\text{ADM}}$ the set ${\cal B}_{i}^{\text{canonical}}$ is indeed the set of fields that need to be fixed at the boundary. As we have mentioned, there is a one-to-one relationship between the continuous fields ${\cal B}_{i}$ required for factorization and the boundary conditions in the variational problem.

In a classical boundary value problem,  the fields in Eq.~(\ref{Continuity_fields_across_T_canonical_action}) represent a minimal set of data that must be fixed in order to have a well-defined functional derivative of the canonical action. However, the classical equations of motion will only admit a solution if the boundary data for $\Pi_{ab}^|$ is consistent with Hamilton's first equation~\eqref{Hamiltons_first_equations}. Relatedly, we have to impose Hamilton's first equation to achieve equivalence with the Lagrangian form of the boundary conditions in Eq.~\eqref{f(Riemann)_Dirichlet_conditions} (using Eq.~\eqref{canonical_fields_radial_projection_definitions} to express \eqref{f(Riemann)_Dirichlet_conditions} in terms of projected fields.) In terms of the projected fields the relevant component reads
\beq
\label{tangential_momenta_Hamiltons}
\frac{\Pi^{|}_{ab}}{\sqrt{h}} = \frac{\sigma}{N}\left( \frac{1}{2}(\pa_{\tau}s_{ab}) - h_{c(a}{^{(2)}}D_{b)}N^{c}_{|} - k_{ab}N_{0}\right)\,.
\eeq
The above equation together with the orthogonality condition $N_{0}|_{\calT}=0$ show that the momenta $\Pi^{|}_{ab}/\sqrt{h}$ depend only on the boundary metric (and tangential derivatives thereof) that is already fixed in the Lagrangian variational problem. The canonical and Lagrangian problems variational problems are consistent with each other when Hamilton's first equation holds. 

Since the boundary metric is fixed in the variational problem, the canonical action in  Eq.~\eqref{ADM_action_1} together with the boundary terms in Eqs.~\eqref{f(Riemann)_boundary_terms_Hamiltonian}-\eqref{f(Riemann)_boundary_terms_Hamiltonian_2} correspond to the canonical ensemble~\cite{Brown:1992br}.\footnote{This justifies the notation ${\cal B}_{i}^{\text{canonical}}$ in Eq.~\eqref{Continuity_fields_across_T_canonical_action}.}
In the next section we utilize the purely bulk form in Eq.~\eqref{Purely_bulk_form_Canonical_action_final_short} to define the microcanonical ensemble for $f$(Riemann) gravity theories and find the set ${\cal B}_{i}^{{\text{micro}}}$.

\subsection{Microcanonical ensemble for $f$(Riemann)}
\label{subsec:Microcanonical ensemble}
The canonical ensemble for $f$(Riemann) corresponds to the canonical action \eqref{ADM_action_1}. The purely bulk form for the canonical ensemble was found in Eq.~\eqref{Purely_bulk_form_Canonical_action_final_short} and the continuity conditions across the factorization surface were found in Eq.~\eqref{Continuity_fields_across_T_canonical_action}. In this sections we derive the microcanonical ensemble~\cite{Brown:1992bq}. 

The prescription basically amounts to adding appropriate boundary terms to the action that change the boundary conditions on ${\cal B}$ from fixed $N$ and shift $N^{a}$ to fixed \emph{energy-surface density} (Brown-York quasilocal energy) $\epsilon_{\text{BY}}$ and \emph{momentum surface-density} $J_{a}$, while all the other boundary conditions remain the same. From the purely bulk form in Eq.~\eqref{Purely_bulk_form_Canonical_action_final_short} we find that the variation among classical solutions of $I_{\text{ADM}}$ is
 \beq
 \label{Variation_on_shell_canonical action}
 \begin{split}
 \delta \left(I_{\text{ADM}}\big |_{\text{on-shell}}\right)&=\int_{\Sigma_{\tau_{1}},\Sigma_{\tau_{2}}}d^{D-1}x\left(p^{\mu\nu}\delta h_{\mu\nu}+ \Pi_{\mu\nu}\delta\Psi^{\mu\nu}\right) \\&
 -\int_{\tau_{1}}^{\tau_{2}}d\tau\int_{\calS_{\tau}}d^{D-2}x\sqrt{s}\left[\epsilon_{\text{BY}} \delta N - J_{a}\delta N^{a} - J^{ab}\delta s_{ab}\right]\\&
+ \int_{\tau_{1}}^{\tau_{2}}d\tau\int_{\calS_{\tau}}d^{D-2}x\sqrt{s}\bigg[2N k_{ab}\,\delta \theta^{ab}_{|}+4N\left(\frac{\Pi^{|}_{a}}{\sqrt{h}}\right)\delta\phi^{a}_{|}+ (D_{\un r}N)\delta\Psi_{0}\bigg]\,,
 \end{split}
 \eeq
 where 
 \beq
 \label{energy_momentum_surface_densities}
 \begin{split}
 \epsilon_{\text{BY}}&=-2\,{^{(2)}}D_{\lambda}\Psi^{\lambda}_{|}+ k_{\alpha\beta}\Psi^{\alpha\beta}_{|} - k\Psi_{0} + 2a_{\lambda}\Psi^{\lambda}_{|} - D_{\un r}\Psi_{0} - 2k_{\alpha\beta}\theta^{\alpha\beta}_{|} - 4\phi^{\lambda}_{|}\frac{\Pi^{|}_{\lambda}}{\sqrt{h}} - 4\phi^{\mu\nu}_{2,|}\left(\Pi^{|}_{\mu\nu}/\sqrt{h}\right)\,,\\
 J_{a}&=2\frac{\Pi_{a\lambda}}{\sqrt{h}}\Psi^{\lambda\nu}r_{\nu}- 2 h_{a\sigma}\frac{p^{\sigma\nu}}{\sqrt{h}}r_{\nu}- 4\sigma\,s_{af}\, {^{(2)}}D_{b}\left(\phi^{(bf)}_{2,|}\right)\,.
 \end{split}
 \eeq
 In the derivation of Eq.~\eqref{Variation_on_shell_canonical action} we used  Hamilton's equation \eqref{tangential_momenta_Hamiltons} for  the term $\phi_{2,|}^{ab}\delta (\Pi^{|}_{ab}/\sqrt{h})$, which is responsible for the final terms in the expressions for $\epsilon_{\text{BY}}$, $J_{a}$ in Eq.~\eqref{energy_momentum_surface_densities}. The precise form of $J^{ab}$ is quite complicated but is not needed for our purposes. Note that in the derivation of \eqref{Variation_on_shell_canonical action} we have assumed the orthogonality condition $N_{0}|_{\calT}=0$ and as usual dropped total derivatives of the form ${^{(2)}}D_{a}v^a$. It is now clear from Eq.~\eqref{Variation_on_shell_canonical action} that in order to define the microcanonical action one needs to add the following boundary terms in the canonical action \eqref{ADM_action_1}
 \beq
 \label{microcanonical_action}
 I_{\text{micro}}=I_{\text{ADM}} + \int_{\tau_{1}}^{\tau_{2}}d\tau \int_{\calS_{\tau}}d^{D-2}x\sqrt{s}\left[N\,\epsilon_{\text{BY}} - J_{a}N^{a} \right]\,.
 \eeq
 From Eq~\eqref{ADM_action_1} and \eqref{f(Riemann)_boundary_terms_Hamiltonian}, \eqref{f(Riemann)_boundary_terms_Hamiltonian_2} the microcanonical action \eqref{microcanonical_action} becomes
\beq
\label{microcanonical_action_explicit_form}
I_{\text{micro}}=\int_{\tau_{1}}^{\tau_{2}}d\tau\int_{\Sigma_{\tau}}d^{D-1}x\left(p^{\mu\nu}\pa_{\tau} h_{\mu\nu}+ \Pi_{\mu\nu}\pa_{\tau}\Psi^{\mu\nu}- N{\cal C}- N^{\mu}{\cal C}_{\mu}\right)+ B_{\text{micro}}\,,
\eeq
where the boundary term $B_{\text{micro}}$ is
\beq
\label{B_micro_boundary_term}
B_{\text{micro}}=-\int_{\tau_{1}}^{\tau_{2}}d\tau\int_{{\cal S}_{\tau}}d^{D-2}x\sqrt{s}\bigg[4N \phi^{ab}_{2,|}\frac{\Pi^{|}_{ab}}{\sqrt{h}}- 4\sigma s_{af}N^{a}\,{^{(2)}}D_{b}\left(\phi^{(bf)}_{2,|}\right) \bigg]\,.
\eeq 
It is interesting to note the appearance of the boundary term above vanishes in GR~\cite{Brown:1992bq} because in that case we have $\phi^	{ab}_{2,|}=0$. 

The variation of $I_{\text{micro}}\big |_{\text{on-shell}}$ given in Eq.~\eqref{microcanonical_action_explicit_form} is 
 \beq
 \begin{split}
 \delta \left(I_{\text{micro}}\big |_{\text{on-shell}}\right) &=\int_{\Sigma_{\tau_{1}},\Sigma_{\tau_{2}}}d^{D-1}x\left(p^{\mu\nu}\delta h_{\mu\nu}+ \Pi_{\mu\nu}\delta\Psi^{\mu\nu}\right) \\&
 + \int_{\tau_{1}}^{\tau_{2}}d\tau\int_{\calS_{\tau}}d^{D-2}x\sqrt{s}\left[N\delta \epsilon_{\text{BY}}  - N^{a} \delta J_{a} - J^{ab}\delta s_{ab}\right]\\&
+ \int_{\tau_{1}}^{\tau_{2}}d\tau\int_{\calS_{\tau}}d^{D-2}x\sqrt{s}\bigg[2N k_{ab}\,\delta \theta^{ab}_{|}+4N\left(\frac{\Pi^{|}_{a}}{\sqrt{h}}\right)\delta\phi^{a}_{|}+ (D_{\un r}N)\delta\Psi_{0}\bigg]\,,
 \end{split}
 \eeq
where we used Eq.~\eqref{Variation_on_shell_canonical action}.

The corresponding microcanonical path integral is
\beq
\label{microcanonical_path_integral}
Z^{\text{micro}}= \int {\cal D} N{\cal D}N^{\mu}{\cal D}h_{\mu\nu}{\cal D}p^{\mu\nu}{\cal D}\Psi^{\mu\nu}{\cal D}\Pi_{\mu\nu}{\cal D}[\Omega_{\mu\nu},\rho_{\mu\nu\kappa\lambda},\rho_{\mu\nu\kappa}\phi^{\mu\nu\kappa\lambda}\phi^{\mu\nu\kappa}]\, e^{- I_{\text{micro}}}\,.
\eeq
The purely bulk form of the microcanonical action in Eq.~\eqref{microcanonical_action_explicit_form} is
\beq
\label{Purely_bulk_form_microcanonical_action_final_short}
\begin{split}
I_{\text{micro}}&=\int_{\tau_{1}}^{\tau_{2}}d\tau \int_{\Sigma_{\tau}}d^{D-1}x \left(p^{\mu\nu}\pa_{\tau}h_{\mu\nu}+ \Pi_{\mu\nu}\pa_{\tau}\Psi^{\mu\nu}\right)\\&
+\int_{\tau_{1}}^{\tau_{2}}d\tau \int_{\Sigma_{\tau}}\sqrt{h}N \blue{D_{\un r}\epsilon_{\text{BY}}} - \int_{\tau_{1}}^{\tau_{2}}d\tau\int_{\Sigma_{\tau}}d^{D-1}x\sqrt{h}N^{a}_{|}\blue{D_{\un r}J_{a}}+ \int_{\tau_{1}}^{\tau_{2}}d\tau\int_{\Sigma_{\tau}}d^{D-1}x\sqrt{h}N\,\blue{k \,\epsilon_{\text{BY}}} \\&
-4\int_{\tau_{1}}^{\tau_{2}}d\tau\int_{\Sigma_{\tau}}d^{D-1}x\sqrt{h}\phi^{\nu\kappa}_{2,|}\blue{D_{\un r} \left(N (\Pi^{|}_{\nu\kappa}/\sqrt{h}) + \sigma {^{(2)}}D_{(\nu}N^{|}_{\kappa)}\right)}\\&
+ 2\int_{\tau_{1}}^{\tau_{2}}d\tau \int_{\Sigma_{\tau}}d^{D-1}x\sqrt{h}N\blue{k_{\mu\nu}D_{\un r}\theta^{\mu\nu}_{|}}
+ 4\int_{\tau_{1}}^{\tau_{2}}d\tau \int_{\Sigma_{\tau}}d^{D-1}x\sqrt{h}N (\blue{D_{\un r}\phi^{\mu}_{|}})\frac{\Pi^{|}_{\mu}}{\sqrt{h}}\\&
+\int_{\tau_{1}}^{\tau_{2}}d\tau \int_{\Sigma_{\tau}}d^{D-1}x\sqrt{h}(\blue{D_{\un r}N_{0}})\left(2r^{\mu}\frac{\Pi_{\mu\lambda}}{\sqrt{h}}\Psi^{\lambda\nu}r_{\nu}- 2r_{\mu}\frac{p^{\mu\nu}}{\sqrt{h}}r_{\nu}\right)\\&
-\int_{\tau_{1}}^{\tau_{2}}d\tau \int_{\Sigma_{\tau}}d^{D-1}x\sqrt{h}N_{0}\bigg(\frac{\Pi^{|}_{\alpha\beta}}{\sqrt{h}}\blue{D_{\un r}\Psi^{\alpha\beta}_{|}}+ 2\frac{\Pi^{|}_{\beta}}{\sqrt{h}}\blue{D_{\un r}\Psi^{\beta}_{|}} + \frac{\Pi_{0}}{\sqrt{h}}\blue{D_{\un r}\Psi_{0}}\bigg) \\&
+ \int_{\tau_{1}}^{\tau_{2}}d\tau \int_{\Sigma_{\tau}}d^{D-1}x\sqrt{h}\,{\cal G'}\left(\blue{k_{\mu\nu}},a_{\mu}, \text{other field combinations with no radial derivatives}\right)\,,
\end{split}
\eeq
where $\epsilon_{\text{BY}}$, $J_{a}$ were defined in \eqref{energy_momentum_surface_densities}. The precise form of ${\cal G'}$ is modified compared to the function ${\cal G}$ in Eq.~\eqref{Purely_bulk_form_Canonical_action_final_short} and its precise form is not needed. Similarly to the canonical ensemble the blue terms in Eq.~\eqref{Purely_bulk_form_microcanonical_action_final_short}  are all the terms that contain radial derivatives. From the purely bulk form of the microcanonical action we find the continuous fields across ${\cal T}$ are
\beq
\label{Continuity_fields_across_T_microcanonical_action}
{\cal B}_{i}^{\text{micro}}= \{\epsilon_{\text{BY}},~ J_{a},~s_{ab},~ \theta^{ab}_{|},~\phi^{a}_{|},~\Psi_{0}\,,~\text{and }~ N(\Pi^{|}_{ab}/\sqrt{h}) + \sigma {^{(2)}}D_{(a}N_{b)}\,\}\quad\mbox{at}~\calT\,,
\eeq
while the lapse $N$ and shift $N^{a}$ can be discontinuous, just as in Einstein gravity. One way to impose continuity of $N(\Pi^{|}_{ab}/\sqrt{h}) + \sigma {^{(2)}}D_{(a}N_{b)}$ would be to impose Hamilton's first equation \eqref{tangential_momenta_Hamiltons} as a surface condition for \eqref{f(Riemann)_Dirichlet_conditions}, upon which continuity of $N(\Pi^{|}_{ab}/\sqrt{h}) + \sigma {^{(2)}}D_{(a}N_{b)}$  becomes continuity of $\frac{1}{2}\sigma\partial_\tau s_{ab} $, which is already continuous according to~(\ref{Continuity_fields_across_T_microcanonical_action}). Since we are discussing factorization, however, which can take place on any surface, this would essentially mean imposing Hamilton's first equation for $\Pi^|_{ab}$ everywhere. The precise continuity condition is somewhat more general.\footnote{A practical approach to imposing it would be to change path integral variables from $\Pi^{|}_{ab}$ to $Z_{ab}$ that are related by $\Pi^{|}_{ab}/\sqrt{h}= N^{-1}Z_{ab}- \sigma {^{(2)}}D_{(a}N_{b)}$, and then enforce continuity of $Z_{ab}$. } 

The fields in Eq.~\eqref{Continuity_fields_across_T_microcanonical_action} were defined in Eqs.~\eqref{canonical_fields_radial_projection_definitions}, \eqref{spatial_fields} and the currents $\epsilon_{\text{BY}}, J_{a}$ in Eq.~\eqref{energy_momentum_surface_densities}. With these continuity properties on the fields the microcanonical action in \eqref{microcanonical_action} satisfies the additivity property \eqref{additivity} and the corresponding path integral \eqref{microcanonical_path_integral} factorizes
\beq
\label{microcanonical_path_integral_f_Riemann}
Z^{\text{micro}}= \int_{\calT} D {\cal B}_{i}^{\text{micro}}\,Z({\calM}_{1}| {\cal B}_{i}^{\text{micro}})Z({\cal B}_{i}^{\text{micro}}| {\calM}_{2})\,.
\eeq

Let us emphasize the role of the discontinuous lapse allowed by the microcanonical path integral. Such a lapse discontinuity is actually realized semiclassically in a number of simple settings involving horizons out of equilibrium. For example, it allows us to make sense of Euclidean $\text{SdS}_{4}$ as a saddle point of a constrained path integral, exhibiting two smooth horizons but a finite lapse discontinuity.  The constraint  amounts to fixing the fields ${\cal B}_{i}^{\text{micro}}$ on the surface ${\cal T}$. We will see this in detail in the next section.
Finally, as we mentioned above the microcanonical ensemble also allows for discontinuous shift $N^{a}$. Although this is not needed for the SdS geometry it is probably important for other geometries, like the (complex Euclidean continuation of) the Kerr-de Sitter spacetime~\cite{Carter:1973rla}.

\section{Constrained states in EFT}
\label{sec:SdS Entropy}
Now we consider a constrained  path integral, obtained by inserting a delta functional to $Z^{\text{micro}}$ that fixes the microcanonical data \eqref{Continuity_fields_across_T_microcanonical_action} to the values $\underbar{{\cal B}}_{i}^{\text{micro}}$ on a codimension-1 surface ${\calT}$. The constrained  path integral then becomes 
\beq \label{constrainedfactorization}
Z^{\text{constrained}}=Z({\calM}_{1}| \underbar{{\cal B}}_{i}^{\text{micro}})Z(\underbar{{\cal B}}_{i}^{\text{micro}}| {\calM}_{2})\,,
\eeq 
i.e., Eq.~(\ref{microcanonical_path_integral_f_Riemann}) with the integral omitted. Similar expressions hold for other types of surface constraints. 

The saddle point approximation may then by applied to each of the two factors separately. Since the boundary data ${\cal B}_{i}$ does not generally include all the field data, the semiclassical solution may be discontinuous on $\calM$. However, the continuity requirements are inferred from a classical gravitational action, which is scale dependent when coupled to quantum fields. For this reason it is important to establish the continuity conditions implied by general Wilsonian effective actions.

The canonical and microcanonical factorization properties of the leading truncation of the effective action, Einstein gravity, were discussed in~\cite{Draper:2022xzl}. Here we have shown that some of the most physically interesting properties -- namely the lack of continuity restrictions on $N$ and $N^a$ --  are robust against ultraviolet quantum corrections, in the sense that the more general effective actions studied in the previous section have the same properties. In particular, a low-curvature, discontinuous-$N$ or -$N^a$ saddle point solution to a constrained path integral in  Einstein gravity, or to a constrained path integral with a more general Wilsonian gravitational effective action, should provide a good approximation to a saddle point solution with similar discontinuities arising from a new gravitational effective action obtained by a small change in renormalization scale.

(Since we have only worked with the $f$(Riemann) class of effective actions, which includes Einstein gravity but not all possible higher dimension operators, and moreover since the local effective action does not incorporate all infrared quantum effects, we cannot draw a completely general conclusion about quantum corrections to solutions of Einstein gravity. However, as discussed in the introduction, $f$(Riemann) is sufficiently general to discuss factorization provided one performs suitable field and coupling redefinitions, at least through the first few orders in the derivative expansion.) 

Furthermore, we may use the Euclidean effective action to compute quantum corrections, parameterized by the effective couplings, to the thermodynamic potentials of constrained states. Examples include the  free energies of out-of-equilibrium black holes in asymptotically flat or AdS space at finite temperature, and the Schwarzschild de Sitter black hole, which is always a constrained state and always out of equilibrium. Let us work out the latter in $D=4$, including general curvature-squared terms in the effective action.

We have a path integral over metrics on a manifold of  $S^{4}$ topology, and a microcanonical constraint applied on a surface corresponding to a fixed ``radial" coordinate. The path integral then factorizes into the form~(\ref{constrainedfactorization}).

We can now apply the saddle point approximation on each factor separately. What are values $\underbar{{\cal B}}_{i}^{\text{micro}}$ should we fix on ${\calT}$? With malice aforethought we will fix them to be the SdS boundary data. To find their precise form, first note that including operators up to dimension 4, the function $f$(Riemann) in Eq.~\eqref{DSSY_action_1} takes the form 
\beq
\label{Riemann_squared_EFT}
f({\cal R}\indices{^{MN}_{K\Lambda}})=-\frac{1}{8\pi G_{N}}\left({\cal R} -2\Lambda\right) + b_{1}\; {\cal R}_{MNK\Lambda}{\cal R}^{MNK\Lambda}+ b_{2}\;{\cal R}_{MN}{\cal R}^{MN}+ b_{3}\;{\cal R}^{2}\,,
\eeq
with $\Lambda>0$. The Einstein-Hilbert case $b_{1}=b_{2}=b_{3}=0$ was analyzed in~\cite{Draper:2022xzl}.
The auxiliary field Lagrangian $I_{\cal L}$ corresponding to the above $f$(Riemann) action is given in Eq.~\eqref{DSSY_action_2} with
\beq
\label{f_Riemann_squared}
f(\varrho\indices{^{MN}_{K\Lambda}})=-\frac{1}{8\pi G_{N}}\left(\varrho -2\Lambda\right) + b_{1}\; \varrho_{MNK\Lambda}\varrho^{MNK\Lambda}+ b_{2}\;\varrho_{MN}\varrho^{MN}+ b_{3}\;\varrho^{2}\,,
\eeq
where $\varrho=g^{MP}g^{N\Sigma}\varrho_{MNP\Sigma}$. 

The last two equations of motion in Eq.~\eqref{eom_g_varphi_varrho} become 
\beq
\label{varphi_varrho_on_shell_Riemann_squared}
\begin{split}
\varphi^{MNK\Lambda} &= -\frac{1}{16\pi G_{N}}\left(g^{MK}g^{N\Lambda}- g^{NK}g^{M\Lambda}\right) + 2b_{1}{\cal R}^{MNK\Lambda}+ b_{3}{\cal R}\left(g^{MK}g^{N\Lambda}- g^{NK}g^{M\Lambda}\right)\\&~~
+  \frac{1}{2}b_{2}\left({\cal R}^{N\Lambda}g^{MK}- {\cal R}^{M\Lambda}g^{NK}-{\cal R}^{NK}g^{M\Lambda}+ {\cal R}^{MK}g^{N\Lambda}\right)\,,\\
\varrho_{MNK\Lambda}&={\cal R}_{MNK\Lambda}\,.
\end{split}
\eeq
Substituting the above into the first equation of motion in \eqref{eom_g_varphi_varrho}, $E^{MN}=0$, we find the following metric equation of motion for the EFT Lagrangian in \eqref{Riemann_squared_EFT},
\beq
\label{metric_eom_Riemann_squared}
\begin{split}
&-\frac{1}{8\pi G_{N}}\left(G^{MN}+ \Lambda g^{MN}\right)- \frac{1}{2}g^{MN}\bigg((b_{2}+4b_{1}){\cal R}_{AB}{\cal R}^{AB}+ (b_{3}-b_{1}){\cal R}^{2}- (b_{2}+ 4b_{3})\nabla_{K}\nabla^{K}{\cal R}\bigg)\\&
- 2(b_{2}+ 4b_{1}){\cal R}_{AB}{\cal R}^{MABN}+ 2(b_{3}-b_{1}){\cal R}{\cal R}^{MN}
- (b_{2}+ 2b_{1}+ 2b_{3})\nabla^{M}\nabla^{N}{\cal R}\\&
 + (4b_{1}+b_{2})\nabla^{K}\nabla_{K}{\cal R}^{MN} =0\,.
\end{split}
\eeq 
It is now easy to check that general Einstein manifolds satisfy the equation of motion if we substitute $R_{MN}= \Lambda g_{MN}$ into Eq.~\eqref{metric_eom_Riemann_squared}. Thus SdS is a solution of the gravitational EFT up to curvature-squared operators. To find the constrained values of $\underbar{{\cal B}}_{i}^{\text{micro}}$ we need to compute them for the SdS geometry. Before we present the result let us describe the SdS geometry in more detail. 

The Euclidean SdS geometry is
\beq
\label{SdS_geometry}
ds^2 = {\cal N}^2 f(\rho)d\tau^2 + f(\rho)^{-1}d\rho^2 + \rho^2 d\Omega^2\,,~~f(\rho)=1 - 2M/\rho - \rho^2/L^2~\,,
\eeq 
where $L=\sqrt{3/\Lambda}$ is the de Sitter radius, $M$ the black hole mass parameter and $d\Omega^2$ the $S_{2}$ metric element. We impose Euclidean time periodicity
 $\tau\sim \tau +\beta$ with $\beta= 1/T_{b}$. We also define the lapse function ${\cal N}=1+ (T_{b}/T_{c}-1)\theta(\rho- r)$ that keeps track of the lapse discontinuity. $T_{b}= \frac{L^2 -4 r_{b}^{2}}{4\pi L^2 r_{b}}$ is the temperature of the black hole horizon located at $\rho=r_{b}$, and $T_{c}= \frac{3 r_{c}^{2}-L^2}{4\pi L^2 r_{c}}$ is the temperature of the cosmological horizon at $\rho=r_{c}$. The factorization surface ${\cal T}_{r}$ is located at radial coordinate $\rho=r$. The Heaviside step function that appears in lapse function ${\cal N}$ indicates that for the black hole side $\rho<r$, we have ${\cal N}=1$, while for the cosmological side $\rho>r$, we have ${\cal N}= \frac{T_{b}}{T_{c}}$. Note that although both sides have the same Euclidean time periodicity $\beta=1/T_{b}$, both horizons are smooth due to the different lapse functions on each side. 
 
 The continuation of the discontinuous Euclidean solution \eqref{SdS_geometry} to Lorentzian singnature can be done as follows \cite{Draper:2022xzl}. First we rescale the Euclidean time periodicity of the cosmological side such that the lapse is continuous but the time periodicity is not. We then continue  $\tau\to t$ on both sides which erases the discontinuity in the periodicity by decompactifying $t$ in both regions. The end result is the Lorentzian SdS solution with mass M.
 
 Let us compute the factorization surface values    $\underbar{{\cal B}}_{i}^{\text{micro}}$ \eqref{Continuity_fields_across_T_microcanonical_action} of the continuous fields for the SdS geometry given in Eq.~\eqref{SdS_geometry}. We find
\beq
\label{SdS_values_B_i_microcanonical}
\begin{split}
\epsilon_{\text{BY}}\big |_{\text{SdS}}&=-2r^{-1}f(r)^{1/2}\left(-\frac{1}{8\pi G_{N}}+ 2\Lambda (4b_{3}+ b_{2})\right) + 2b_{1}f(r)^{1/2}\left(2r^{-1}f''(r) +f'''(r)\right)\\
J_{a} \big |_{\text{SdS}}&=0~~,~~s_{ab}\big |_{\text{SdS}}=r^{2}t_{ab}~,~
\Psi_{0}\big |_{\text{SdS}}=-\frac{1}{8\pi G_{N}}+ 2 \Lambda \left(4b_{3}+b_{2}\right)- 2b_{1}f''(r)\\
\theta^{ab}_{|}|_{\text{SdS}}&= r^{-2}\bigg[\left(-\frac{1}{16\pi G_{N}}+ \Lambda (4b_{3}+b_{2})\right) - b_{1} r^{-1}f'(r)\bigg]t^{ab}~,~\phi^{a}_{|}\big |_{\text{SdS}}=0\\
N(\Pi^{|}_{ab}&/\sqrt{h})+  {^{(2)}}D_{(a}N_{b)}\big |_{\text{SdS}}=0\,,
\end{split}
\eeq
where $t_{ab}$ are the unit round metric components in $(\theta,\varphi)$ coordinates and $f(\rho)$ is given in Eq.~\eqref{SdS_geometry}. 
The details of the proof of Eqs.~\eqref{SdS_values_B_i_microcanonical} are presented in Appendix \ref{subsec:Calculation of SdS microcanonical data}. The constrained path integral \eqref{constrainedfactorization} with $\underbar{{\cal B}}_{i}^{\text{micro}}$ fixed to \eqref{SdS_values_B_i_microcanonical}  on $\calT$ will have the SdS geometry \eqref{SdS_geometry} on both sides of $\calT$ as a saddle point solution. Then in the leading saddle approximation, if there are no saddles of lower action, $Z^{\text{constrained}} \approx Z^{\text{constrained}}_{\text{SdS}}$. The geometry is a genuine saddle because the geometry solves the equations of motion everywhere off of the constraint surface -- in particular, both horizons are smooth. 

Thus SdS can be interpreted as a constrained state of the gravitational path integral with an effective action up through arbitrary curvature-squared operators. Beyond this order in the derivative expansion of the effective action, the saddle point geometry will receive corrections from the effective couplings. Our results from the previous section show that if the corrections are of $f$(Riemann) type then there is no obstruction to maintaining a discontinuous lapse function, so we expect that the change in the saddle point geometry will be a small effect, well controlled in the derivative and semiclassical expansions. 

Finally, let us compute the curvature squared corrections to the entropy of the constrained SdS state. We need to compute the on-shell value of the microcanonical action \eqref{microcanonical_action_explicit_form}. Although the momenta $p^{\mu\nu},\Pi_{\mu\nu}$, time derivatives $\pa_{\tau}h_{\mu\nu},\,\pa_{\tau}\Psi^{\mu\nu}$ and the boundary terms in \eqref{microcanonical_action_explicit_form} vanish for SdS, one cannot conclude that the on-shell action vanishes. The reason is that the time foliation breaks down at the horizons $r_{b,c}$. To calculate the on-shell action we will use the method of~\cite{Banados:1993qp} which we  refer to as the \emph{BTZ prescription}. 

The BTZ prescription is as follows: first we draw infinitesimal boundaries $\calT_{b}$ and $\calT_{c}$ around the black hole and cosmological horizons. To compute the on-shell action we  use the ADM form of the microcanonical action for the region between $\calT_{b}$ and $\calT_{c}$, while for the regions inside the boundaries $\calT_{b}$ and $\calT_{c}$ we  use the Lagrangian form of the microcanonical action. This means that we  use the microcanonical action \eqref{microcanonical_action} in these infinitesimal regions, but instead of $I_{\text{ADM}}$, we  use the Lagrangian form \eqref{DSSY_action_4}. For the large bulk region in between $\calT_{b}$ and $\calT_{c}$, the microcanonical action vanishes on-shell for SdS. To see this note that on-shell the lapse and shift constraints should hold ${\cal C}={\cal C}_{\mu}=0$ and since $\pa_{\tau}h_{\mu\nu},\pa_{\tau}\Psi^{\mu\nu}$ vanish for SdS, the bulk terms in Eq.~\eqref{microcanonical_action_explicit_form} also vanish. The boundary terms $B_{\text{micro}}$ in Eq.~\eqref{B_micro_boundary_term} also vanish because $\Pi^{|}_{ab}|_{\text{SdS}}= N^{a}|_{\text{SdS}}=0$. For the infinitesimal horizon regions ``inside" $\calT_{b}$ and $\calT_{c}$, one can show that the boundary terms in Eq.~\eqref{microcanonical_action} vanish for the infinitesimal surfaces $\calT_{b,c}$. The bulk terms in \eqref{DSSY_action_4} also vanish because the curvature is finite but the integration volume goes to zero. The entirety of the classical action is thus due to generalized GHY terms ${\cal S}^{{\cal B}}_{\text{GHY}}$ on $\calT_{b,c}$. From Eq.~\eqref{Generalized_GHY_term} and the SdS metric \eqref{SdS_geometry}, we find
\beq
\label{microcanonical_action_SdS_entropy}
\begin{split}
 I_{\text{micro}}^{\text{SdS}}&=- \frac{\pi r_{b}^{2}+ \pi r_{c}^{2}}{G_{N}} + 64\pi^{2}b_{1}+ 48\pi^2 \frac{r_{b}^{2}+ r_{c}^{2}}{L^2}(4b_{3}+ b_{2})\\&
 = -\frac{A_{b}+ A_{c}}{4G_{N}} + 64\pi^{2}b_{1}+ 12\pi(4b_{3}+b_{2}) \frac{A_{b}+ A_{c}}{L^{2}}\,,
 \end{split}
\eeq
where the horizons $r_{b},r_{c}$ satisfy $f(r_{b,c})=0$ with $f(\rho)$ given in \eqref{SdS_geometry} and $A_{b,c}=4\pi r_{b,c}^{2}$. 
We see that the result is independent of the location $r$ where the constrained was imposed. The entropy is minus the total on-shell action. 

We have verified by explicit computation that the same result for the entropy is obtained using Wald's formula~\cite{Wald:1993nt, Jacobson:1993vj}. Namely, we find 
\beq 
\label{Wald_entropy_SdS}
\begin{split}
S^{\text{SdS}}_{\text{W}}&:= -2\pi \oint_{\text{horzn.}}d\theta d\varphi\sqrt{s}\, \hat\epsilon^{MN}\hat\epsilon_{K\Lambda}\frac{\pa {\cal L}}{\pa {\cal R}\indices{^{MN}_{K\Lambda}}}\\&\,\,
= -  I_{\text{micro}}^{\text{SdS}}
\end{split}
\eeq
where ${\cal L}= -\frac{1}{2}f({\cal R}\indices{^{MN}_{K\Lambda}})$ \footnote{The minus sign in ${\cal L}=-\frac{1}{2}f({\cal R}\indices{_{MNK\Lambda}})$ is due to rotation from Euclidean to Lorentzian signature.} and $f({\cal R}\indices{^{MN}_{K\Lambda}})$ was given in Eq.~\eqref{Riemann_squared_EFT}, $s_{ab}= r^{2}t_{ab}$ is the cod-2 induced metric at the horizon, $\hat\epsilon_{MN}= \nabla_{M}\tilde \xi_{N}$ and $\tilde\xi^{M}$ is the Killing field that is null at the horizon with normalized surface gravity $\kappa:=-\frac{1}{2}\nabla_{A}\tilde\xi_{B}\nabla^{A}\tilde\xi^{B}\big |_{\text{horzn.}}=1$. Some details of the computation are presented at the end of Appendix \ref{subsec:Calculation of SdS microcanonical data}. 

Let us make some remarks on the relation between BTZ and Wald's method of calculating the entropy. The relation with the ordinary Gibbons-Hawking Euclidean approach was already pointed by Wald \cite{Wald:1993nt}. The equivalence of the former to the BTZ method then implies that we have two methods, one Lorentzian and one Euclidean, for computing horizon entropies just from data localized near the horizons. Thus there should be a geometric demonstration that the two computations are equivalent for geometries with Killing horizons. Let us assume spherical topology for the Euclidean horizon (or bifurcate horizon in Lorentzian signature). Then the BTZ surface $\calT_{r}$ enclosing the  Euclidean horizon $r_{h}$ is $S_{\beta_{r}}^{1}\times S^{D-2}$, where $\beta_{r}$ denotes the size of $S_{\beta_{r}}^{1}$. In the horizon limit $\beta_{r}$ shrinks to zero and the horizon has finite area. For $f$(Riemann) Lagrangians, we can compare Wald's formula in Eq.~\eqref{Wald_entropy_SdS} with the generalized  GHY term in Eqs.~\eqref{Generalized_GHY_term},\eqref{DSSY_K_Psi_GHY}, and using the on-shell value for $\varphi_{MNK\Lambda}$ given in Eq.~\eqref{eom_g_varphi_varrho} we find that for the two formulas to match we must have 
\beq
\label{Wald_is_BTZ}
\lim_{r\to r_{h}}~\beta_{r}\, r_{[M}{\cal K}_{A][B}\,r_{N]}=\lim_{r\to r_{h}}\left( -\frac{\pi}{2}\nabla_{M}\tilde\xi_{A}\nabla_{B}\tilde\xi_{N}\right)\,,
\eeq
where the brackets on the left hand side mean anti-symmetrization and the right hand side is computed using the appropriate Euclidean continuation of the Killing vector $\un{\tilde \xi}$. 
We note that  both sides of Eq.~\eqref{Wald_is_BTZ} should be computed in a  coordinate system that is well-defined at the Euclidean horizon, e.g. Kruskal-Szekeres coordinates, where the Riemann components are finite.
It is easy to check directly that Eq.~\eqref{Wald_is_BTZ} holds for the Schwarzschild horizon.

Returning to SdS, finally we can estimate that the nucleation probability of black hole of mass M in the de Sitter background is
\beq
P= Z^{\text{constrained}}/Z\sim e^{S_{\text{SdS}}- S_{\text{dS}}}= e^{ -I_{\text{micro}}^{\text{SdS}}- \frac{\pi L^2}{G_{N}} + 16\pi^2 \left(4b_{1}+ 12b_{3}+ 3b_{2}\right)}\,,
\eeq
where $I_{\text{micro}}^{\text{SdS}}$ was given in \eqref{microcanonical_action_SdS_entropy}.

\section{Conclusions}
\label{sec:Conclusions}

Let us summarize our main results and sketch some possible directions for future progress.

The primary general result is the derivation of the continuity properties of the gravitational fields in  factorizations of the path integral, corresponding to Dirichlet and microcanonical factorizations, in a large class of gravitational effective Lagrangians. These continuity properties can be used to implement the semiclassical approximation
with fixed data on a factorization surface, including the insertion of constraints. 

Our main application of the general result was to the Euclidean Schwarzschild-de Sitter geometry, where the detailed continuity properties of the factorized path integral are already necessary to make semiclassical sense of the geometry in Einstein gravity. By showing that the continuity properties of Einstein gravity extend in a simple way to a large class of more general effective actions, we conclude that the discontinuity present in Euclidean Schwarzschild-de Sitter is not an accident of working in the leading truncation of the effective Lagrangian, and something SdS-like can be expected to persist after taking into account the quantum fluctuations of massive matter fields. At the next-to-leading truncation, SdS itself remains a solution, and the BTZ method of computing the on-shell action -- which reduces to a horizon-localized integral in the Euclidean signature -- matches the Lorentzian signature result obtained by Wald's formula.

Let us now discuss some possible extensions of this work.

As discussed above, we consider effective actions of the $f$(Riemann) type. This restricts the EFT operators to the parity-even sector with no explicit covariant derivatives. (In some cases, or at least through low orders, it is possible to perform field redefinitions and integrations-by-parts to rewrite the action in the  $f$(Riemann) form, but we are not aware of a general proof that this can be extended to arbitrary order.) At a fixed order in the derivative expansion, we believe it should be possible to incorporate other classes of terms by introducing a finite number of additional auxiliary fields, but it would be interesting to exhibit this construction explicitly.

In addition, as  briefly mentioned at the end of Section \ref{sec:Path Integral Factorization}, one could investigate other types of factorizations apart from fixing canonical or microcanonical data. For example, one could fix the surface angular momentum density $J_a$ along with $\epsilon_{\text{BY}}$ to non-zero values.  It would be interesting to examine whether the Kerr-de Sitter metrics, which are complex metrics in Euclidean signature,  provide a consistent semiclassical approximation of the contribution to the partition function.

Finally,  we have confined our analysis to the leading order in the semiclassical expansion. Going beyond the leading order introduces many interesting technical and conceptual complications. Some have been discussed in Sec. \ref{subsec:Local Symmetries}. Issues of ellipticity that affect ordinary one-sided boundaries in gravity~\cite{Witten:2018lgb} are not a problem for factorization surfaces (essentially because they are two-sided, and ``boundary moving diffs" should be treated as genuine diffs), but they do highlight the technical issue of gauge fixing in the presence of a factorization surface. Another interesting related question concerns the renormalization of boundary terms, and thus of semiclassical boundary conditions: are all relevant boundary conditions radiatively stable? 

We hope to address these matters in future work. Nearly a half century after the discovery of the Gibbons-Hawking entropy, the gravitational path integral continues to pose as many questions as it answers.

~\\

\section*{Acknowledgements} 
This work was supported in part by the U.S.
Department of Energy, Office of Science, Office of High Energy Physics under award number DE-SC0015655.

~\\

\appendix

\section{Reduction to General Relativity}
\label{subsec:General Relativity}
In this appendix we reduce the path integral with the auxiliary fields to the gravitational path integral of pure General Relativity (GR). There are two different approaches for showing this, based on the Lagrangian or canonical path integral. We will present both.\footnote{We note that our manipulations do not keep track of determinants that should be present when integrating out fields but do not affect leading semiclassics.}

In the first approach we use the Lagrangian formalism defined by the action \eqref{DSSY_action_2} and path integral \eqref{DSSY_path_integral} and reduce to the GR Lagrangian by integrating out the auxiliary fields. We find the effective Lagrangian \eqref{eff_ADM_action_GR}  is indeed the EH Lagrangian with the GHY boundary term. Similarly we repeat the procedure starting from the canonical form of the action \eqref{ADM_action_1} and path integral \eqref{f(Riemann)_ADM_path_integral}. In this second approach we find the system indeed reduces to the usual GR canonical action \eqref{eff_canonical_ADM_action_GR_v2} with GHY boundary term. We will present both approaches.
\newline
 For GR the function $f(\varrho_{MNP\Sigma})$ appearing in \eqref{DSSY_action_2} is 
\beq\label{f_rho_GR}
f^{\text{GR}}(\varrho\indices{^{MN}_{K\Lambda}})= -\frac{\sigma}{8\pi G_{N}}\left(g^{MP}g^{N\Sigma}\varrho_{MNP\Sigma}-2\Lambda\right)\,.
\eeq
Note that in GR given the function $f^{GR}(\varrho\indices{^{MN}_{K\Lambda}})$ defined in \eqref{f_rho_GR} above, the fields $\rho_{MNK\Lambda}$ appear linearly in the action \eqref{DSSY_action_2} and thus act as Lagrange multipliers. Integrating them out along the imaginary contour in field space inserts the Dirac delta functional in the path integral \eqref{DSSY_path_integral}. After integrating out the fields $\varphi^{MNK\Lambda}$, the Dirac delta functional imposes the following constraint
\beq\label{varphi_GR}
\varphi_{\text{GR}}^{MNK\Lambda}= -\frac{\sigma}{16\pi G_{N}}\left(g^{MK}g^{N\Lambda}- g^{NK}g^{M\Lambda}\right)\,.
\eeq 
Enforcing the constraint \eqref{varphi_GR} to the generalized GHY term \eqref{Generalized_GHY_term} we obtain the usual GHY boundary term of GR,
\beq\label{GHY_GR_case}
{\cal S}^{\text{GR}}_{\text{GHY}}= \frac{\sigma}{8\pi G_{N}}\epsilon \int_{\pa {\calM}}d^{D-1}y  \sqrt{|r|}{\cal K}\,,
\eeq
where ${\cal K}:= g^{MN}{\cal K}_{MN}$ is the extrinsic curvature trace. In deriving \eqref{GHY_GR_case} we used ${\cal \Psi}_{\text{GR}}^{MN}= \frac{\sigma}{8\pi G_{N}}\epsilon\, r^{MN}$ which can be proven from the definition \eqref{DSSY_K_Psi_GHY} if we use the constraint  \eqref{varphi_GR}. The effective action for the metric degrees of freedom can be found by integrating out $\varrho_{MNP\Sigma}$ and then $\varphi^{MNK\Lambda}$ in the partition function \eqref{DSSY_action_2}
\beq\label{Z_effective_lagrangian}
Z= \int {\cal D} g_{MN}\, e^{-I^{\text{eff}}_{{\cal L}}[g_{MN}]}\,.
\eeq
Enforcing the constraint \eqref{varphi_GR}, we find the effective action $I^{\text{eff}}_{{\cal L}}[g_{MN}]$ to be the Einstein-Hilbert action with  GHY boundary term
\beq\label{eff_ADM_action_GR}
I^{\text{eff}}_{{\cal L}}[g_{MN}]= -\frac{\sigma}{16\pi G_{N}}\int_{{\calM}}d^{D}x \sqrt{|g|} \left({\cal R}-2\Lambda\right) -\frac{\sigma}{8\pi G_{N}} \epsilon \int_{\pa {\calM}} \sqrt{|r|}d^{D-1}y {\cal K}\,.
\eeq
For Euclidean metrics we have $\sigma=\epsilon=1$ which matches the action in equation (2) of~\cite{Draper:2022xzl}.\footnote{The GR path integral \eqref{Z_effective_lagrangian} can be expressed for both Lorentzian/ Euclidean signature $\sigma=\mp1$ as $Z= \int \left[{\cal D} g_{MN}\right] e^{-\sigma \sqrt{\sigma}I^{\text{eff}}_{{\cal L}}[g_{MN}]}$, where $i=\sqrt{-1}$ and $I^{\text{eff}}_{{\cal L}}[g_{MN}]$ is given in \eqref{eff_ADM_action_GR}.}

We now repeat the same procedure, starting from the ADM form of the action \eqref{ADM_action_1} with corresponding path integral \eqref{f(Riemann)_ADM_path_integral}. The function $f^{\text{GR}}(\varrho_{MNK\Lambda})$ in \eqref{f_rho_GR} can be expressed in terms of the spatial fields
\beq
\label{f_rho_GR_canonical}
f^{\text{GR}}(\varrho\indices{^{MN}_{K\Lambda}})=-\frac{\sigma}{8\pi G_{N}}\left(h^{\mu\lambda}h^{\nu\sigma} \rho_{\mu\nu\lambda\sigma} + 2\sigma h^{\mu\nu}\Omega_{\mu\nu}- 2\Lambda\right)\,,
\eeq
where $h_{\mu\nu}$ is the induced metric on $\Sigma_{\tau}$, and the fields $\rho_{\mu\nu\lambda\sigma}$, $\Omega_{\mu\nu}$ are defined in \eqref{spatial_fields}. Note that from  \eqref{f_rho_GR_canonical} and the lapse constraint ${\cal C}$ given in \eqref{boundary_terms_lapse_shift_constraints} the fields $\Omega_{\mu\nu}$, $\rho_{\mu\nu\kappa\lambda}$, $\rho_{\mu\nu\kappa}$ appear linearly in the ADM action \eqref{ADM_action_1}. Integrating them out along the imaginary contour inserts the following delta functionals to the path integral\footnote{We inserted appropriate factors $1/\sqrt{h}$ in the delta functionals using the inner product convention $(A,B)=\int  d^{D-1}x \sqrt{h}\,A\cdot B$ for tensors $A$,$ B$.}
\beq
\label{delta_functionals_canonical}
\begin{split}
&\delta \left(\sigma N\left(\Psi^{\mu\nu} + \frac{\sigma}{8\pi G_{N}}h^{\mu\nu}\right)\right)~,~\delta \left(\frac{N}{2}\left(\frac{\sigma}{16\pi G_{N}}(h^{\mu\kappa}h^{\nu\lambda}-h^{\nu\kappa}h^{\mu\lambda}) + \phi^{\mu\nu\kappa\lambda}\right)\right)\,,\\&
\delta \left(2\sigma N\,\phi^{\mu\nu\kappa}\right)\,.
\end{split}
\eeq
If we integrate out $\Psi^{\mu\nu}$, $\phi^{\mu\nu\kappa\lambda}$, $\phi^{\mu\nu\kappa}$ the delta functional arguments are imposed as constraints and reduce the path integral \eqref{f(Riemann)_ADM_path_integral} to 
\beq
\label{effective_path_integral_1}
Z= \int {\cal D} N{\cal D}N^{\mu}{\cal D}h_{\mu\nu}{\cal D}p^{\mu\nu}{\cal D}\Pi_{\mu\nu}\, e^{- \tilde{I}_{\text{ADM}}}\,,
\eeq
where
\beq
\tilde{I}_{\text{ADM}} = \int_{\tau_{1}}^{\tau_{2}}d\tau\int_{\Sigma_{\tau}}d^{D-1}x\left((p^{\mu\nu}+ \frac{\sigma}{8\pi G_{N}}\Pi^{\mu\nu})\pa_{\tau}h_{\mu\nu}- N \tilde{\cal C}- N^{\mu}\tilde {\cal C}_{\mu}\right)+\tilde B_{N} + \tilde B_{\un N} \,, 
\eeq
and $\tilde{\cal C}$, $\tilde {\cal C}_{\mu}$, $\tilde B_{N}$, $\tilde B_{\un N}$ are
\beq
\label{constraints_tilde_GR_appendix}
\begin{split}
\tilde{\cal C} &=\frac{1}{\sqrt{h}}\frac{1}{16\pi G_{N}}\left(3\Pi\cdot\Pi + \Pi^{2}\right) +  \frac{\sigma}{16\pi G_{N}}\sqrt{h}\left(R- 2\Lambda\right) + \frac{1}{\sqrt{h}}2\sigma p\cdot\Pi \,,\\
\tilde {\cal C}_{\mu}&=-\frac{\sigma }{8\pi G_{N}}\sqrt{h} h_{\mu\lambda}D_{\nu}\left(2\frac{\Pi\indices{^{\lambda\nu}}}{\sqrt{h}}\right)- 2\sqrt{h}h_{\mu\lambda}D_{\nu}\left(\frac{p^{\lambda\nu}}{\sqrt{h}}\right)\,, \\
\tilde B_{N}&=-\frac{\sigma}{8\pi G_{N}}\int_{\tau_{1}}^{\tau_{2}}d\tau\int_{{\cal S}_{\tau}}d^{D-2}x\sqrt{|s|}Nk\,,\\
\tilde B_{\un N}&=-2\int_{\tau_{1}}^{\tau_{2}}d\tau\int_{\calS_{\tau}}d^{D-2}x\sqrt{s}N^{\mu} h_{\mu\sigma}\left(\frac{p^{\sigma\nu}}{\sqrt{h}}+ \frac{\sigma}{8\pi G_{N}}\frac{\Pi\indices{^{\sigma\nu}}}{\sqrt{h}}\right)r_{\nu}\,.
\end{split}
\eeq
Note that there is one extra degree of freedom left in the path integral \eqref{effective_path_integral_1} compared to the GR canonical path integral. One way to proceed is to integrate out $p^{\mu\nu}$, which yields the GR path integral in Lagrangian form. Let us take a different route and obtain it in the phase space form. We define the new momenta $\ti p^{\mu\nu},\ti \Pi^{\mu\nu}$
\beq
\label{rotated_momenta_tilde_GR_v2}
\ti p^{\mu\nu} = p^{\mu\nu}+ \frac{\sigma}{8\pi G_{N}}\Pi^{\mu\nu}~~,~~\ti \Pi^{\mu\nu}=p^{\mu\nu}- \frac{\sigma}{8\pi G_{N}}\Pi^{\mu\nu}\,,
\eeq
after which the $\tilde I_{\text{ADM}}$ takes the more familiar form
\beq
\label{ADM_action_constraint_plus_one_dof_GR_v2}
\begin{split}
\tilde I_{\text{ADM}}&=\int_{\tau_{1}}^{\tau_{2}}d\tau\int_{\Sigma_{\tau}}d^{D-1}x\left(\ti p^{\mu\nu}\pa_{\tau}h_{\mu\nu}- N\tilde{\cal C} - N^{\mu}\tilde {\cal C}_{\mu}\right)+\tilde B_{N}+ \tilde B_{\un N}\,,
\end{split}
\eeq
where $\tilde{\cal C}, \tilde {\cal C}_{\mu}$ and the boundary terms $\tilde B_{\un N}$,$\tilde B_{N}$ of Eq.~\eqref{constraints_tilde_GR_appendix} expressed in terms of $\ti p^{\mu\nu}$ and $\ti \Pi^{\mu\nu}$ are
\beq
\begin{split}
\tilde{\cal C} &=\frac{\pi G_{N}}{\sqrt{h}}\left(7\ti p\cdot\ti p - 6\ti p\cdot\ti\Pi -2\ti p\ti \Pi -\ti \Pi\cdot\ti\Pi+ \ti p^{2} + \ti \Pi^2 \right) + \frac{\sigma}{16\pi G_{N}}\sqrt{h}\left(R-2\Lambda\right)\,,\\
\tilde {\cal C}_{\mu}&= -2 \sqrt{h}h_{\mu\lambda}D_{\nu}\left(\frac{\ti p^{\lambda\nu}}{\sqrt{h}}\right)\,, \\
\tilde B_{N}&=-\frac{\sigma}{8\pi G_{N}}\int_{\tau_{1}}^{\tau_{2}}d\tau\int_{{\cal S}_{\tau}}d^{D-2}x\sqrt{s}Nk\,,\\
\tilde B_{\un N}&=-2\int_{\tau_{1}}^{\tau_{2}}d\tau\int_{\calS_{\tau}}d^{D-2}x\sqrt{s}N^{\mu} h_{\mu\sigma}\left(\frac{\tilde p^{\sigma\nu}}{\sqrt{h}}\right)r_{\nu}\,.
\end{split}
\eeq
Note that $\ti \Pi^{\mu\nu}$ appears quadratically in \eqref{ADM_action_constraint_plus_one_dof_GR_v2}. Integrating out $\ti \Pi^{\mu\nu}$ is equivalent to substituting the  $\ti \Pi^{\mu\nu}$ variation equations of motion
\beq\label{on_shell_tilde_Pi_GR_v2}
\ti \Pi^{\mu\nu}=- 3\ti p^{\mu\nu} + \frac{4}{D-2}\ti p h^{\mu\nu}~~,~~\ti \Pi= \frac{D+2}{D-2}\ti p\,,
\eeq
into \eqref{ADM_action_constraint_plus_one_dof_GR_v2}. The path integral \eqref{effective_path_integral_1} then reduces to the GR canonical path integral 
\beq
\label{path_integral_GR_ADM_effective_action_v2}
Z= \int {\cal D} N\,{\cal D} N^{\mu}\,{\cal D} h_{\mu\nu}\,{\cal D}\ti p^{\mu\nu}\, e^{-I^{\text{eff}}_{\text{ADM}}}\,,
\eeq
where
\beq
\label{eff_canonical_ADM_action_GR_v2}
\begin{split}
I^{\text{eff}}_{\text{ADM}}=&\int_{\tau_{1}}^{\tau_{2}}d\tau\int_{\Sigma_{\tau}}d^{D-1}x\left(\ti p^{\mu\nu}\pa_{\tau}h_{\mu\nu}- N\tilde{\cal C}_{\text{eff}}- N^{\mu}\tilde{\cal C}^{\text{eff}}_{\mu}\right)+\tilde B_{N}+ \tilde B_{\un N}\,,
\end{split}
\eeq
and
\beq
\label{lapse_shift_GR_effective_v2}
\begin{split}
\tilde {\cal C}_{\text{eff}}&= \frac{16\pi G_{N}}{\sqrt{h}}\left(\ti p^{\mu\nu}\ti p_{\mu\nu} -\frac{1}{D-2}\ti p^2 \right) + \frac{\sigma}{16\pi G_{N}}\sqrt{h}\left(R-2\Lambda\right)\,,\\
\tilde{\cal C}^{\text{eff}}_{\mu}&=-2 \sqrt{h}h_{\mu\lambda}D_{\nu}\left(\frac{\ti p^{\lambda\nu}}{\sqrt{h}}\right)\,, \\
\tilde B_{N}&=-\frac{\sigma}{8\pi G_{N}}\int_{\tau_{1}}^{\tau_{2}}d\tau\int_{{\cal S}_{\tau}}d^{D-2}x\sqrt{s}Nk\,,\\
\tilde B_{\un N}&=-2\int_{\tau_{1}}^{\tau_{2}}d\tau\int_{\calS_{\tau}}d^{D-2}x\sqrt{s}N^{\mu} h_{\mu\lambda}\left(\frac{\tilde p^{\lambda\nu}}{\sqrt{h}}\right)r_{\nu}\,.
\end{split}
\eeq
The action \eqref{eff_canonical_ADM_action_GR_v2} together with the lapse and shift constraints \eqref{lapse_shift_GR_effective_v2} is the well known canonical form of the EH action with the GHY boundary term in D dimensions.

\section{ Derivation of purely bulk form of $I_{\text{ADM}}$}
\label{subsec:Derivation of purely bulk form}
In this appendix we derive the complete form of the purely bulk form given in Eq.~\eqref{Purely_bulk_form_Canonical_action_final_short}. 
We start with the canonical form of the action given in Eq.~\eqref{ADM_action_1}. By definition the purely bulk form should have no boundary terms so our goal is to cancel $B_{N}$, $B_{\un N}$ given in Eqs.~\eqref{f(Riemann)_boundary_terms_Hamiltonian}-\eqref{f(Riemann)_boundary_terms_Hamiltonian_2}. As we will see shortly, some of the bulk terms after integration by parts will generate boundary terms that cancel $B_{N}$, $B_{\un N}$.

The first step is to integrate by parts the first two terms in the shift constraint ${\cal C}_{\mu}$ in Eq.~\eqref{boundary_terms_lapse_shift_constraints}. The covariant derivative for these two terms will get shifted to $N^{\mu}$ and the boundary term generated will cancel the $B_{\un N}$ boundary term. The next step is to integrate by parts the first term of the lapse constraint ${\cal C}$ in Eq.~\eqref{boundary_terms_lapse_shift_constraints}. The boundary term generated will cancel the first term in $B_{N}$ given in Eq.~\eqref{f(Riemann)_boundary_terms_Hamiltonian}. After these two integrations by parts the canonical action \eqref{ADM_action_1} becomes
\beq
\label{Factorization_ADM_action_purely_bulk_1}
\begin{split}
I_{\text{ADM}}&= \int_{\tau_{1}}^{\tau_{2}}d\tau \int_{\Sigma_{\tau}} d^{D-1}x \bigg[p^{\mu\nu}\pa_{\tau}h_{\mu\nu}+ \Pi_{\mu\nu}\pa_{\tau}\Psi^{\mu\nu} \\&
+ \sqrt{h}D_{\nu}N^{\mu}\left(2 \frac{\Pi_{\mu\lambda}}{\sqrt{h}}\Psi^{\lambda\nu}- 2 h_{\mu\sigma}\frac{p^{\sigma\nu}}{\sqrt{h}}\right)
- \sqrt{h} \left(\frac{\Pi_{\alpha\beta}}{\sqrt{h}}N^{\lambda}D_{\lambda}\Psi^{\alpha\beta}\right) + \sqrt{h}(D_{\mu}N)D_{\lambda}\Psi^{\mu\lambda}\\&
+ \sqrt{h}N\left(\frac{1}{2}\phi^{\mu\nu\kappa\lambda}R_{\mu\nu\kappa\lambda}\right) 
+ 4\sqrt{h} N \phi^{\mu\nu\kappa}D_{\mu}\left(\frac{\Pi_{\nu\kappa}}{\sqrt{h}}\right)\\&
- \sqrt{h}N \left(-\frac{1}{2}f(\varrho\indices{^{MN}_{K\Lambda}})+ \sigma \Psi\cdot \Omega+ 2\sigma \sqrt{h}\phi^{\mu\nu\kappa}\rho_{\mu\nu\kappa}\right)\\&
- N \left(\frac{1}{\sqrt{h}}\left(2\sigma\, p^{\mu\nu}\Pi_{\mu\nu}- \sigma\,\Psi\cdot\Pi\cdot\Pi- \sigma\,\Pi (\Psi\cdot\Pi)\right)- \frac{\sqrt{h}}{2}\phi^{\mu\nu\kappa\lambda}\left(\frac{2\sigma}{h}\Pi_{\mu\lambda}\Pi_{\nu\kappa}- \rho_{\mu\nu\kappa\lambda}\right)\right)
\bigg] \\&
+ \int_{\tau_{1}}^{\tau_{2}}d\tau\int_{\calS_{\tau}}d^{D-2}x\sqrt{s}\bigg[- Nr_{\nu}\Psi^{\nu\lambda}D_{\lambda}\ln N + 2Nk_{\mu\nu}\phi^{\mu\gamma\nu\delta}r_{\gamma}r_{\delta}+ 4Nr^{\beta}\left(\frac{\Pi_{\alpha\beta}}{\sqrt{h}}\right)\phi^{\alpha\gamma\delta}r_{\gamma}r_{\delta}\\&\qquad\qquad\qquad\qquad \qquad~ + N(D_{\un r}\ln N)(r_{\alpha}r_{\beta}\Psi^{\alpha\beta})\bigg]\,.
\end{split}
\eeq
Note there are boundary terms remaining in the above expression so the action is not in purely bulk form yet. However the terms in the second and third line of Eq.~\eqref{Factorization_ADM_action_purely_bulk_1} above contain terms which after integration by parts will cancel the remaining boundary terms. To see this explicitly we use Eq.~\eqref{canonical_fields_radial_projection} and find the following expressions 
\beq
\label{GC_canonical_action_terms_appendix}
\begin{split}
D_{\nu}N^{\mu}&\left(2\frac{\Pi_{\mu\lambda}}{\sqrt{h}}\Psi^{\lambda\nu}- 2 h_{\mu\sigma}\frac{p^{\sigma\nu}}{\sqrt{h}}\right)=D_{\un r}N^{\mu}_{|}\left(2\frac{\Pi_{\mu\lambda}}{\sqrt{h}}\Psi^{\lambda\nu}r_{\nu}- 2 h_{\mu\sigma}\frac{p^{\sigma\nu}}{\sqrt{h}}r_{\nu}\right)\\&
+ D_{\un r}N_{0}\left(2r^{\mu}\frac{\Pi_{\mu\lambda}}{\sqrt{h}}\Psi^{\lambda\nu}r_{\nu}- 2r_{\mu}\frac{p^{\mu\nu}}{\sqrt{h}}r_{\nu}\right) \\&
+ \left({^{(2)}}D_{\nu}N^{\mu}_{|}+ (k\indices{^{\mu}_{\nu}}+ r_{\nu}a^{\mu})N_{0}+ r^{\mu}({^{(2)}}D_{\nu}N_{0}- N_{|}^{\sigma}k_{\sigma\nu})\right)\left(2 \frac{\Pi_{\mu\lambda}}{\sqrt{h}}\Psi^{\lambda\nu}- 2 h_{\mu\sigma}\frac{p^{\sigma\nu}}{\sqrt{h}}\right)\\\\
\frac{\Pi_{\alpha\beta}}{\sqrt{h}} N^{\lambda}D_{\lambda}\Psi^{\alpha\beta}&= \frac{\Pi^{|}_{\alpha\beta}}{\sqrt{h}} N^{\lambda}\,{^{(2)}}D_{\lambda}\Psi^{\alpha\beta}_{|} + 2\frac{\Pi^{|}_{\alpha\beta}}{\sqrt{h}} N^{\lambda}_{|}k\indices{_{\lambda}^{\alpha}}\Psi^{\beta}_{|}
- 2\frac{\Pi^{|}_{\beta}}{\sqrt{h}}N_{|}^{\lambda}k\indices{_{\lambda\alpha}}\Psi^{\alpha\beta}_{|} + 2\frac{\Pi^{|}_{\beta}}{\sqrt{h}} N^{\lambda}_{|}{^{(2)}}D_{\lambda}\Psi^{\beta}_{|}\\&
+ 2\frac{\Pi^{|}_{\beta}}{\sqrt{h}} N^{\lambda}_{|}k\indices{_{\lambda}^{\beta}}\Psi_{0} 
- 2\frac{\Pi_{0}}{\sqrt{h}}N^{\lambda}_{|}k\indices{_{\lambda\beta}}\Psi^{\beta}_{|}+ \frac{\Pi_{0}}{\sqrt{h}}N_{|}^{\lambda}{^{(2)}}D_{\lambda}\Psi_{0} - 2\frac{\Pi_{0}}{\sqrt{h}}N_{0} a_{\beta}\Psi^{\beta}_{|}+2 \frac{\Pi^{|}_{\beta}}{\sqrt{h}}N_{0}a^{\beta}\Psi_{0}\\&
+ \frac{\Pi^{|}_{\alpha\beta}}{\sqrt{h}} N_{0}D_{\un r}\Psi_{|}^{\alpha\beta} + 2\frac{\Pi^{|}_{\alpha\beta}}{\sqrt{h}} N_{0}a^{\alpha}\Psi^{\beta}_{|} + 2N_{0}\frac{\Pi^{|}_{\beta}}{\sqrt{h}}D_{\un r}\Psi^{\beta}_{|} +  \frac{\Pi_{0}}{\sqrt{h}}N_{0}D_{\un r}\Psi_{0}- 2N_{0}\frac{\Pi^{|}_{\beta}}{\sqrt{h}}a_{\lambda}\Psi^{\lambda\beta}_{|}\,\\\\
(D_{\mu}N)D_{\lambda}\Psi^{\mu\lambda}&={^{(2)}}D_{\mu}N \left({^{(2)}}D_{\lambda}\Psi^{\mu\lambda}_{|}- \Psi_{|}^{\mu\lambda}a_{\lambda}+ \Psi_{|}^{\sigma}k\indices{_{\sigma}^{\mu}}+ a^{\mu}\Psi_{0}\right)\\&
- (D_{\un r} N)\left(k_{\alpha\beta}\Psi^{\alpha\beta}_{|}- 2{^{(2)}}D_{\lambda}\Psi^{\lambda}_{|} - k \Psi_{0}+ 2\Psi^{\lambda}_{|}a_{\lambda}- D_{\un r}\Psi_{0}\right)\\&
+ D_{\mu}\left(r^{\mu}\Psi_{|}^{\lambda}~{^{(2)}}D_{\lambda}N\right)- D_{\lambda}\left(\Psi^{\lambda}_{|}D_{\un r}N\right)+ \Psi^{\lambda}_{|}k\indices{_{\lambda}^{\sigma}}{^{(2)}}D_{\sigma}N\,,\\\\
\phi^{\mu\nu\kappa\lambda}R_{\mu\nu\kappa\lambda}&=\theta_{|}^{\mu\nu\kappa\lambda}{^{(2)}}R_{\mu\nu\kappa\lambda}+ \theta_{|}^{\mu\nu\kappa\lambda}\left(k_{\mu\lambda}k_{\nu\kappa}-k_{\mu\kappa}k_{\nu\lambda}\right)+ 4 \theta_{|}^{\mu\nu\lambda}({^{(2)}}D_{\mu}k_{\lambda\nu}- {^{(2)}}D_{\nu}k_{\lambda\mu})\\&
+ 4\theta_{|}^{\mu\nu}(- k\indices{_{\mu}^{\sigma}}k_{\sigma\nu}+ {^{(2)}}D_{\mu}a_{\nu}- a_{\mu}a_{\nu})+ 4k \theta^{\mu\nu}_{|}k_{\mu\nu}+ 4k_{\mu\nu}D_{\un r}\theta_{|}^{\mu\nu} - 4 D_{\mu}\left(r^{\mu}\theta_{|}^{\alpha\beta}k_{\alpha\beta}\right)\,,\\\\
\phi^{\mu\nu\kappa}D_{\mu}\left(\frac{\Pi_{\nu\kappa}}{\sqrt{h}}\right)&=\phi^{\mu\nu\kappa}_{|}{^{(2)}}D_{\mu}\left(\frac{\Pi^{|}_{\nu\kappa}}{\sqrt{h}}\right) + \phi^{\mu\nu\kappa}_{|}k_{\mu\kappa}\frac{\Pi^{|}_{\nu}}{\sqrt{h}} -\phi^{\mu\kappa}_{2,|}k\indices{_{\mu}^{\sigma}}\frac{\Pi^{|}_{\sigma\kappa}}{\sqrt{h}}+ \phi^{\mu\kappa}_{2,|}{^{(2)}}D_{\mu}\left(\frac{\Pi^{|}_{\kappa}}{\sqrt{h}}\right)\\&
- 2\phi_{2,|}^{\mu\nu}\frac{a_{(\mu}\Pi^{|}_{\nu)}}{\sqrt{h}}
 - \phi_{1,|}^{\mu\nu}k\indices{_{\mu}^{\kappa}}\frac{\Pi^{|}_{\nu\kappa}}{\sqrt{h}}+ \phi_{1,|}^{\mu\nu}\,{^{(2)}}D_{\mu}\left(\frac{\Pi^{|}_{\nu}}{\sqrt{h}}\right)+ \phi^{\nu}_{|}a^{\mu}\frac{\Pi^{|}_{\mu\nu}}{\sqrt{h}}\\& - 2\phi^{\mu}_{|}k\indices{_{\mu}^{\sigma}}\frac{\Pi^{|}_{\sigma}}{\sqrt{h}}+ \phi_{|}^{\mu}\,{^{(2)}}D_{\mu}\left(\frac{\Pi_{0}}{\sqrt{h}}\right)- \phi^{\mu}_{|}a_{\mu}\frac{\Pi_{0}}{\sqrt{h}}+ k\phi^{\mu}_{|}\frac{\Pi^{|}_{\mu}}{\sqrt{h}}\\&
-D_{\mu}\left(r^{\mu}\phi^{\lambda}_{|}\frac{\Pi^{|}_{\lambda}}{\sqrt{h}}\right)+ \frac{\Pi^{|}_{\mu}}{\sqrt{h}}D_{\un r}\phi_{|}^{\mu} -\phi^{\nu\kappa}_{2,|}D_{\un r}\left(\frac{\Pi^{|}_{\nu\kappa}}{\sqrt{h}}\right)\,.
\end{split}
\eeq
For the definition of the projected fields (with $|$ subscript) see Eq.~\eqref{canonical_fields_radial_projection_definitions}. If we substitute expressions \eqref{GC_canonical_action_terms_appendix} into Eq.~\eqref{Factorization_ADM_action_purely_bulk_1} the boundary terms cancel from the total derivative terms in the last three equations of Eq.~\eqref{GC_canonical_action_terms_appendix}, and we finally find the purely bulk form of the canonical action
\beq
\label{Factorization_ADM_action_purely_bulk_complete_form}
\begin{split}
&I_{\text{ADM}}=\int_{\tau_{1}}^{\tau_{2}}d\tau \int_{\Sigma_{\tau}}d^{D-1}x \left(p^{\mu\nu}\pa_{\tau}h_{\mu\nu}+ \Pi_{\mu\nu}\pa_{\tau}\Psi^{\mu\nu}\right)\\&
+\int_{\tau_{1}}^{\tau_{2}}d\tau \int_{\Sigma_{\tau}}d^{D-1}x \sqrt{h}\blue{(D_{\un r} N)}\left[2\,{^{(2)}}D_{\lambda}\Psi^{\lambda}_{|}-\blue{k_{\alpha\beta}}\Psi^{\alpha\beta}_{|} + \blue{k} \Psi_{0}- 2a_{\lambda}\Psi^{\lambda}_{|}+ \blue{D_{\un r}\Psi_{0}} + 2\blue{k_{\alpha\beta}}\theta^{\alpha\beta}_{|} + 4\phi^{\lambda}_{|}\frac{\Pi^{|}_{\lambda}}{\sqrt{h}}\right]\\&
+\int_{\tau_{1}}^{\tau_{2}}d\tau \int_{\Sigma_{\tau}}d^{D-1}x\sqrt{h}N\left[(\blue{D_{\un r}N^{\mu}_{|}})\left(2\frac{\Pi_{\mu\lambda}}{\sqrt{h}}\Psi^{\lambda\nu}r_{\nu}- 2 h_{\mu\sigma}\frac{p^{\sigma\nu}}{\sqrt{h}}r_{\nu}\right)+ \blue{D_{\un r}N_{0}}\left(2r^{\mu}\frac{\Pi_{\mu\lambda}}{\sqrt{h}}\Psi^{\lambda\nu}r_{\nu}- 2r_{\mu}\frac{p^{\mu\nu}}{\sqrt{h}}r_{\nu}\right) \right]\\&
- 4\int_{\tau_{1}}^{\tau_{2}}d\tau \int_{\Sigma_{\tau}}d^{D-1}x\sqrt{h}N\phi^{\nu\kappa}_{2,|}\blue{D_{\un r}\left(\frac{\Pi^{|}_{\nu\kappa}}{\sqrt{h}}\right)}+ 2\int_{\tau_{1}}^{\tau_{2}}d\tau \int_{\Sigma_{\tau}}d^{D-1}x\sqrt{h}N\blue{k_{\mu\nu}D_{\un r}\theta^{\mu\nu}_{|}}\\&
 + 4\int_{\tau_{1}}^{\tau_{2}}d\tau \int_{\Sigma_{\tau}}d^{D-1}x\sqrt{h}N (\blue{D_{\un r}\phi^{\mu}_{|}})\frac{\Pi^{|}_{\mu}}{\sqrt{h}}\\&
-\int_{\tau_{1}}^{\tau_{2}}d\tau \int_{\Sigma_{\tau}}d^{D-1}x\sqrt{h}N_{0}\left(\frac{\Pi^{|}_{\alpha\beta}}{\sqrt{h}}\blue{D_{\un r}\Psi^{\alpha\beta}_{|}}+ 2\frac{\Pi^{|}_{\beta}}{\sqrt{h}}\blue{D_{\un r}\Psi^{\beta}_{|}} + \frac{\Pi_{0}}{\sqrt{h}}\blue{D_{\un r}\Psi_{0}}\right)\\&
+ \int_{\tau_{1}}^{\tau_{2}}d\tau \int_{\Sigma_{\tau}}d^{D-1}x\sqrt{h}\left({^{(2)}}D_{\nu}N^{\mu}_{|}+ (\blue{k\indices{^{\mu}_{\nu}}}+ r_{\nu}a^{\mu})N_{0}+ r^{\mu}({^{(2)}}D_{\nu}N_{0}- N_{|}^{\sigma}\blue{k_{\sigma\nu}})\right)\left(2 \frac{\Pi_{\mu\lambda}}{\sqrt{h}}\Psi^{\lambda\nu}- 2 h_{\mu\sigma}\frac{p^{\sigma\nu}}{\sqrt{h}}\right)\\&
-\int_{\tau_{1}}^{\tau_{2}}d\tau \int_{\Sigma_{\tau}}d^{D-1}x\sqrt{h}\bigg(\frac{\Pi^{|}_{\alpha\beta}}{\sqrt{h}} N^{\lambda}\,{^{(2)}}D_{\lambda}\Psi^{\alpha\beta}_{|} + 2\frac{\Pi^{|}_{\alpha\beta}}{\sqrt{h}} N^{\lambda}_{|}\blue{k\indices{_{\lambda}^{\alpha}}}\Psi^{\beta}_{|}
- 2\frac{\Pi^{|}_{\beta}}{\sqrt{h}}N_{|}^{\lambda}\blue{k\indices{_{\lambda\alpha}}}\Psi^{\alpha\beta}_{|} + 2\frac{\Pi^{|}_{\beta}}{\sqrt{h}} N^{\lambda}_{|}{^{(2)}}D_{\lambda}\Psi^{\beta}_{|}\\&
+ 2\frac{\Pi^{|}_{\beta}}{\sqrt{h}} N^{\lambda}_{|}\blue{k\indices{_{\lambda}^{\beta}}}\Psi_{0} 
- 2\frac{\Pi_{0}}{\sqrt{h}}N^{\lambda}_{|}\blue{k\indices{_{\lambda\beta}}}\Psi^{\beta}_{|}+ \frac{\Pi_{0}}{\sqrt{h}}N_{|}^{\lambda}{^{(2)}}D_{\lambda}\Psi_{0} - 2\frac{\Pi_{0}}{\sqrt{h}}N_{0} a_{\beta}\Psi^{\beta}_{|}+2 \frac{\Pi^{|}_{\beta}}{\sqrt{h}}N_{0}a^{\beta}\Psi_{0}\\&
+ 2\frac{\Pi^{|}_{\alpha\beta}}{\sqrt{h}} N_{0}a^{\alpha}\Psi^{\beta}_{|} - 2N_{0}\frac{\Pi^{|}_{\beta}}{\sqrt{h}}a_{\lambda}\Psi^{\lambda\beta}_{|}\bigg)\\&
+\int_{\tau_{1}}^{\tau_{2}}d\tau \int_{\Sigma_{\tau}}d^{D-1}x\sqrt{h}{^{(2)}}D_{\mu}N \left({^{(2)}}D_{\lambda}\Psi^{\mu\lambda}_{|}- \Psi_{|}^{\mu\lambda}a_{\lambda}+ 2\Psi_{|}^{\sigma}\blue{k\indices{_{\sigma}^{\mu}}}+ a^{\mu}\Psi_{0}\right)\\&
+\frac{1}{2}\int_{\tau_{1}}^{\tau_{2}}d\tau \int_{\Sigma_{\tau}}d^{D-1}x \sqrt{h} \bigg(\theta_{|}^{\mu\nu\kappa\lambda}{^{(2)}}R_{\mu\nu\kappa\lambda}+ \theta_{|}^{\mu\nu\kappa\lambda}\left(\blue{k_{\mu\lambda}k_{\nu\kappa}}-\blue{k_{\mu\kappa}k_{\nu\lambda}}\right)\\&
+ 4 \theta_{|}^{\mu\nu\lambda}({^{(2)}}D_{\mu}\blue{k_{\lambda\nu}} - {^{(2)}}D_{\nu}\blue{k_{\lambda\mu})}
+ 4\theta_{|}^{\mu\nu}(-\blue{k\indices{_{\mu}^{\sigma}}k_{\sigma\nu}}+ {^{(2)}}D_{\mu}a_{\nu}- a_{\mu}a_{\nu})+ 4\blue{k} \theta^{\mu\nu}_{|}\blue{k_{\mu\nu}}\bigg)\\&
+4\int_{\tau_{1}}^{\tau_{2}}d\tau \int_{\Sigma_{\tau}}d^{D-1}x\sqrt{h}N\bigg[\phi^{\mu\nu\kappa}_{|}{^{(2)}}D_{\mu}\left(\frac{\Pi^{|}_{\nu\kappa}}{\sqrt{h}}\right) + \phi^{\mu\nu\kappa}_{|}\blue{k_{\mu\kappa}}\frac{\Pi^{|}_{\nu}}{\sqrt{h}} -\phi^{\mu\kappa}_{2,|}\blue{k\indices{_{\mu}^{\sigma}}}\frac{\Pi^{|}_{\sigma\kappa}}{\sqrt{h}}+ \phi^{\mu\kappa}_{2,|}{^{(2)}}D_{\mu}\left(\frac{\Pi^{|}_{\kappa}}{\sqrt{h}}\right)\\&
- 2\phi_{2,|}^{\mu\nu}\frac{a_{(\mu}\Pi^{|}_{\nu)}}{\sqrt{h}}
 - \phi_{1,|}^{\mu\nu}\blue{k\indices{_{\mu}^{\kappa}}}\frac{\Pi^{|}_{\nu\kappa}}{\sqrt{h}}+ \phi_{1,|}^{\mu\nu}\,{^{(2)}}D_{\mu}\left(\frac{\Pi^{|}_{\nu}}{\sqrt{h}}\right)+ \phi^{\nu}_{|}a^{\mu}\frac{\Pi^{|}_{\mu\nu}}{\sqrt{h}} - 2\phi^{\mu}_{|}\blue{k\indices{_{\mu}^{\sigma}}}\frac{\Pi^{|}_{\sigma}}{\sqrt{h}}+ \phi_{|}^{\mu}\,{^{(2)}}D_{\mu}\left(\frac{\Pi_{0}}{\sqrt{h}}\right)\\&
 - \phi^{\mu}_{|}a_{\mu}\frac{\Pi_{0}}{\sqrt{h}}+ \blue{k}\phi^{\mu}_{|}\frac{\Pi^{|}_{\mu}}{\sqrt{h}}\bigg]
 - \int_{\tau_{1}}^{\tau_{2}}d\tau \int_{\Sigma_{\tau}}d^{D-1}x\sqrt{h}N \left(-\frac{1}{2}f(\varrho\indices{^{MN}_{K\Lambda}})+ \sigma \Psi\cdot \Omega+ 2\sigma \sqrt{h}\phi^{\mu\nu\kappa}\rho_{\mu\nu\kappa}\right)\\&
- \int_{\tau_{1}}^{\tau_{2}}d\tau \int_{\Sigma_{\tau}}d^{D-1}x\sqrt{h}N \left(\frac{1}{h}\left(2\sigma\, p\cdot\Pi - \sigma\,\Psi\cdot\Pi\cdot\Pi- \sigma\,\Pi (\Psi\cdot\Pi)\right)- \frac{1}{2}\phi^{\mu\nu\kappa\lambda}\left(\frac{2\sigma}{h}\Pi_{\mu\lambda}\Pi_{\nu\kappa}- \rho_{\mu\nu\kappa\lambda}\right)\right)\,.
\end{split}
\eeq
All the terms in Eq.~\eqref{Factorization_ADM_action_purely_bulk_complete_form} that contain radial derivatives are colored with blue.

\section{SdS entropy/microcanonical data}
\label{subsec:Calculation of SdS microcanonical data}
In this appendix we present details of the computation of the microcanonical boundary data $\underbar{{\cal B}}_{i}^{\text{micro}}$ in Eqs.~\eqref{SdS_values_B_i_microcanonical} and SdS entropy from Wald's formula given in Eq.~\eqref{Wald_entropy_SdS}.

We are going to use some geometric facts about the SdS geometry given in Eq.~\eqref{SdS_geometry}
\beq
\label{SdS_geometry_appendix}
ds^2 = {\cal N}^2 f(\rho)d\tau^2 + f(\rho)^{-1}d\rho^2 + \rho^2 d\Omega^2\,,~~f(\rho)=1 - 2M/\rho - \rho^2/L^2~\,,
\eeq 
where $d\Omega^{2}=d\theta^{2}+ \text{sin}^{2}\theta d\varphi^{2}:= t_{ab}dx^{a}dx^{b}$ and $x^{a}=\{\theta,\varphi\}$ are the usual polar coordinates on the sphere. We note that the quantities computed hold for points except at $\rho=r$ where the lapse is  discontinuous.\footnote{For the continuous quantities that we will compute, the value at $\rho=r$ can be defined by taking the limit.} The normal unit form $n$ and vector $\un n$ defined in Eq.~\eqref{normal vector_tau} in $(\tau,\rho,\theta,\varphi)$ coordinates are
\beq
\label{time_unit_normal_SdS_appendix}
n \big |_{\text{SdS}}= {\cal N}f(\rho)^{1/2} d\rho~~,~~\un n\big |_{\text{SdS}}= {\cal N}^{-1}f(\rho)^{-1/2}\un\pa_{\tau}\,.
\eeq
The induced metric \eqref{ADM_cod_2_purely_bulk} on $\Sigma_{\tau}$ time slices in $x^{\mu}=\{\rho,\theta,\varphi\}$ is
\beq
h_{\mu\nu}dx^{\mu}dx^{\nu}=f(\rho)^{-1}d\rho^2 + \rho^2 t_{ab}dx^{a}dx^{b}\,.
\eeq
The radial unit one form $r$ and vector $\un r$ defined in Eq.~\eqref{radial_unit_form} are $r|_{\text{SdS}} = f(\rho)^{-1/2}d\rho~, ~\un r |_{\text{SdS}}= f(\rho)^{1/2}\un\pa_{\rho}$, the radial acceleration $a_{\mu}\big |_{\text{SdS}}=0$, and the radial extrinsic curvature tensor \eqref{radial_extrinsic_tensor} components $k_{\mu\nu}$ are
\beq
\begin{split}
k_{ab}\big |_{\text{SdS}}&= \rho f(\rho)^{1/2} t_{ab} ~~,~~k_{\rho \rho}=k_{\rho a}=0\,,\\
k\big |_{\text{SdS}}:&= h^{ab}k_{ab}= 2\rho^{-1}f(\rho)^{1/2}\,.
\end{split}
\eeq 
It is convenient to use the Gauss Codazzi equations \eqref{Gauss_Codazzi_Ricci}, \eqref{Gauss_Codazzi_Ricci_cod_2_v2}and the fact that for SdS $K_{MN}=0$ to find the following Riemann tensor components for SdS
\beq
\label{Riemann_components_diagonal_f(rho)_appendix}
\begin{split}
{\cal R}_{\rho \tau \rho \tau}\big |_{\text{SdS}}&=- \frac{1}{2}{\cal N}^{2}f''(\rho)~,~{\cal R}_{a \tau b \tau}\big |_{\text{SdS}} =-\frac{1}{2}{\cal N}^{2}\rho f(\rho) f'(\rho)t_{ab}~,~{\cal R}_{\mu\nu\lambda\tau}\big |_{\text{SdS}}=0\,,\\
{\cal R}_{a\rho b\rho}\big |_{\text{SdS}}&=-\frac{\rho}{2}\frac{f'(\rho)}{f(\rho)}t_{ab}~,~
{\cal R}_{abc\rho}\big |_{\text{SdS}}=0~,~
{\cal R}_{abcd}\big |_{\text{SdS}}=\rho^2 (f(\rho)-1)(t_{ad}t_{cb}- t_{ac}t_{bd})\,.
\end{split}
\eeq
The formulas above will be useful for computing the boundary data $\underbar{{\cal B}}_{i}^{\text{micro}}$. First we compute $\phi^{MNK\Lambda}$ for SdS using the formula in Eq.~\eqref{varphi_varrho_on_shell_Riemann_squared}. We find 
 \beq
 \label{varphi_on_SdS_type}
 \varphi^{MNK\Lambda}\big |_{\text{SdS}_{4}}=\left(-\frac{1}{16\pi G_{N}} + \Lambda (4b_{3}+ b_{2})\right)\left(g^{MK}g^{N\Lambda}- g^{NK}g^{M\Lambda}\right)\big |_{\text{SdS}}+ 2b_{1}{\cal R}^{MNK\Lambda}\big |_{\text{SdS}}\,,
 \eeq
 where $\Lambda= 3/L^{2}$.
We use the above formula \eqref{varphi_on_SdS_type} to compute (some) of the spatial fields defined in Eq.~\eqref{spatial_fields} for SdS using the formulas \eqref{Riemann_components_diagonal_f(rho)_appendix}. 
We find
\beq
\label{spatial_fields_SdS_appendix}
\begin{split}
\phi^{a\rho b\rho}\big |_{\text{SdS}}&=\rho^{-2}f(\rho)\bigg[\left(-\frac{1}{16\pi G_{N}}+ \Lambda (4b_{3}+b_{2})\right) - b_{1}\rho^{-1}f'(\rho)\bigg]t^{ab}\,,\\
\phi^{abc\rho}\big |_{\text{SdS}}&=0\,,\\
\phi^{abcd}\big |_{\text{SdS}}&=\rho^{-4}\bigg[\left(-\frac{1}{16\pi G_{N}}+ \Lambda (4b_{3}+b_{2})\right) - 2 b_{1}\rho^{-2} (f(\rho)-1)\bigg](t^{ac}t^{bd}- t^{bc}t^{ad})\,,\\
\phi^{abcd}\big |_{\text{SdS}}&=\rho^{-4}\bigg[\left(-\frac{1}{16\pi G_{N}}+ \Lambda (4b_{3}+b_{2})\right) - 2 b_{1}\rho^{-2} (f(\rho)-1)\bigg](t^{ac}t^{bd}- t^{bc}t^{ad})\,,\\
\phi^{\mu\nu\rho}\big |_{\text{SdS}}&=0\,,\\
\Psi^{ab}\big |_{\text{SdS}}&=2\rho^{-2}\bigg[-\frac{1}{16\pi G_{N}} + \Lambda (4b_{3}+ b_{2}) - b_{1}\rho^{-1}f'(\rho)\bigg]t^{ab}\,,\\
\Psi^{\rho a}\big |_{\text{SdS}}&=0\,,\\
\Psi^{\rho\rho}\big |_{\text{SdS}}&=2 f(\rho)\left(-\frac{1}{16\pi G_{N}}+ \Lambda (4b_{3}+b_{2})- b_{1}f''(\rho)\right)\,.
\end{split}
\eeq
Having found the spatial fields components above we can compute the radial projected field defined in \eqref{canonical_fields_radial_projection_definitions} using \eqref{spatial_fields_SdS_appendix},
\beq
\label{radially_projected_spatial_fields_SdS_appendix}
\begin{split}
N_{|}^{\mu}\bigg |_{\text{SdS}}&= 0~~,~~N_{0}\bigg |_{\text{SdS}}= 0\,, \\
\Pi^{|}_{\mu\nu}\bigg |_{\text{SdS}}&= 0~~,~~ \Pi^{|}_{\mu}\bigg |_{\text{SdS}}= 0~~,~~ \Pi_{0}\bigg |_{\text{SdS}} = 0\,,\\
\Psi_{|}^{ab}\bigg |_{\text{SdS}}&=2\rho^{-2}\bigg[-\frac{1}{16\pi G_{N}} + \Lambda (4b_{3}+ b_{2}) - b_{1}\rho^{-1}f'(\rho)\bigg]t^{ab}~~,~~ \Psi_{|}^{\mu}\bigg |_{\text{SdS}}= 0~~,\\ \Psi_{0}\bigg |_{\text{SdS}}&=-\frac{1}{8\pi G_{N}}+ 2 \Lambda \left(4b_{3}+b_{2}\right)- 2b_{1}f''(\rho)\,,~~~D_{\un r}\Psi_{0}=-2b_{1}f(\rho)^{1/2}f'''(\rho)\,,\\
\theta^{abcd}_{|}|_{\text{SdS}}&=\rho^{-4}\bigg[\left(-\frac{1}{16\pi G_{N}}+ \Lambda (4b_{3}+b_{2})\right) - 2 b_{1}\rho^{-2} (f(\rho)-1)\bigg](t^{ac}t^{bd}- t^{bc}t^{ad})~,~\theta_{|}^{abc}|_{\text{SdS}}=0\,,\\ 
\theta^{ab}_{|}|_{\text{SdS}}&= \rho^{-2}\bigg[\left(-\frac{1}{16\pi G_{N}}+ \Lambda (4b_{3}+b_{2})\right) - b_{1}\rho^{-1}f'(\rho)\bigg]t^{ab}\,, \\
\phi^{abc}_{|}|_{\text{SdS}}&=0,~\phi^{ab}_{1,|}|_{\text{SdS}}=0,~\phi^{ab}_{2,|}|_{\text{SdS}}= 0,~\phi_{|}^{a}|_{\text{SdS}}= 0\,.
\end{split}
\eeq
We can now compute the $\underbar{{\cal B}}_{i}^{\text{micro}}$, namely the fields in Eq.~\eqref{Continuity_fields_across_T_microcanonical_action} for $\text{SdS}_{4}$, namely
\beq
\label{B_micro_SdS_appendix}
\begin{split}
\epsilon_{\text{BY}}\big |_{\text{SdS}}&=-2\rho^{-1}f(\rho)^{1/2}\left(-\frac{1}{8\pi G_{N}}+ 2\Lambda (4b_{3}+ b_{2})\right) + 2b_{1}f(\rho)^{1/2}\left(2\rho^{-1}f''(\rho) +f'''(\rho)\right)\,,\\
J_{a} \big |_{\text{SdS}}&=0~~,~~s_{ab}\big |_{\text{SdS}}=\rho^{2}t_{ab}~,~
\Psi_{0}\big |_{\text{SdS}}=-\frac{1}{8\pi G_{N}}+ 2 \Lambda \left(4b_{3}+b_{2}\right)- 2b_{1}f''(\rho)\,,\\
\theta^{ab}_{|}|_{\text{SdS}}&= \rho^{-2}\bigg[\left(-\frac{1}{16\pi G_{N}}+ \Lambda (4b_{3}+b_{2})\right) - b_{1}\rho^{-1}f'(\rho)\bigg]t^{ab}~,~\phi^{a}_{|}\big |_{\text{SdS}}=0\,,\\
N(\Pi^{|}_{ab}/&\sqrt{h})+  {^{(2)}}D_{(a}N_{b)}\big |_{\text{SdS}}=0\,.
\end{split}
\eeq
Evaluating the formulas in Eq.~\eqref{B_micro_SdS_appendix} at the location $\rho=r$ of the factorization surface ${\cal T}_{r}$ we arrive at Eq.~\eqref{SdS_values_B_i_microcanonical} given in Section \ref{sec:SdS Entropy}.

We end this appendix by presenting some details of the calculation SdS entropy from Wald's formula given in Eq.~\eqref{Wald_entropy_SdS}. The first thing to note is that Wal'ds formulas uses Lorentzian signature metrics. This amounts to rotating ${\cal N}=i$ and decompactifying the time coordinate $\tau\to t$ in Eq.~\eqref{SdS_geometry}. For $f(Riemann)$ type theories the formula reduces to Eq.~\eqref{Wald_entropy_SdS}. The minus sign in ${\cal L}= -(1/2) f({\cal R}\indices{^{MN}_{K\Lambda}})$ also comes from the continuation to Lorentzian signature. There will be two contributions to the entropy coming from the two killing horizons for the Lorentzian SdS geometry. If we use Eq.~\eqref{eom_g_varphi_varrho}  and its value for SdS~\eqref{varphi_on_SdS_type} into the first line of \eqref{Wald_entropy_SdS} for each horizon $\rho= r_{H}$ we get
\beq 
\label{S_Wald_r_H}
\begin{split}
S^{\text{SdS}}_{\text{W}}\big |_{r_{H}}&= -2\pi \oint_{\text{horzn.}}d\theta d\varphi\sqrt{s}\, \hat\epsilon^{MN}\hat\epsilon_{K\Lambda}\frac{\pa {\cal L}}{\pa {\cal R}\indices{^{MN}_{K\Lambda}}}\\&
=\pi  \oint_{\text{horzn.}}d\theta d\varphi \sqrt{s}\hat\epsilon^{MN}\hat\epsilon_{K\Lambda}\,\varphi\indices{_{MN}^{K\Lambda}}\\&
=\pi  \oint_{\text{horzn.}}d\theta d\varphi \sqrt{s}\left[\left(-\frac{1}{16\pi G_{N}} + \Lambda (4b_{3}+ b_{2})\right) 2\hat\epsilon^{MN}\hat\epsilon_{MN} + 2b_{1}\hat\epsilon^{MN}\hat\epsilon^{K\Lambda}{\cal R}^{\text{SdS}}_{MNK\Lambda}\right] \\&
=\pi r_{H}^{2} - 48\pi ^2 \frac{r_{H}^{2}}{L^2} (4b_{3}+ b_{2}) + 8b_{1}\pi^2 r_{H}^{2}\hat\epsilon^{MN}\hat\epsilon^{K\Lambda}{\cal R}^{\text{SdS}}_{MNK\Lambda}
\end{split}
\eeq
where we used $\Lambda=3/L^2$, $\hat\epsilon^{MN}\hat\epsilon_{MN}=-2$, since $\un{\tilde\xi}$ is normalized to unit surface gravity. The quantity $\hat\epsilon^{MN}\hat\epsilon^{K\Lambda}{\cal R}^{\text{SdS}}_{MNK\Lambda}$ can computed at the static coordinate system at some $\rho=r$ and then take the horizon limit $r\to r_{H}$. Being a scalar quantity we expect to get the correct value at horizon by taking the limit even though the static coordinates break down at the horizon. We find using \eqref{Riemann_components_diagonal_f(rho)_appendix} that $\lim_{\rho\to r_{H}}\hat\epsilon^{MN}\hat\epsilon^{K\Lambda}{\cal R}^{\text{SdS}}_{MNK\Lambda}=4{\cal R}^{\text{SdS}}_{\tau\rho\tau\rho}|_{r_{H}}= - 4/r_{H}^{2}$, and if we substitute to the last line of Eq.~\eqref{S_Wald_r_H} we finally find
\beq
\begin{split}
S^{\text{SdS}}_{\text{W}}\big |_{r_{H}}&= \pi r^{2}_{H} - 48\pi^{2}\frac{r_{H}^{2}}{L^2}(4b_{3}+b_{2})- 32 \pi^{2}b_{1}\,.
\end{split}
\eeq
If we add the results for both black hole and cosmological horizons $r_{H}=r_{b,c}$ we find the result  presented in Eqs.~\eqref{microcanonical_action_SdS_entropy}-\eqref{Wald_entropy_SdS}.

~\\

\bibliography{F_Riemann_manuscript_Arxiv}{}
\bibliographystyle{utphys}

\end{document}